\documentclass[prd,twocolumn,amsmath,amssymb,floatfix,superscriptaddress,showpacs]{revtex4}
\usepackage{mathrsfs}
\usepackage{graphicx}
\usepackage{color}
\usepackage{dcolumn}
\usepackage{bm}
\usepackage{float}
\usepackage{subfigure}
\usepackage{multirow}

\allowdisplaybreaks

\begin{document}

\title{Gravitational wave source localization for eccentric binary coalesce with a ground-based detector network}

\author{Sizheng Ma}
\affiliation{Department of Physics and Center for Astrophysics, Tsinghua University, Haidian District, Beijing 100084, China}
\author{Zhoujian Cao}
\email{zjcao@amt.ac.cn}
\affiliation{Department of Astronomy, Beijing Normal University, Beijing 100875, China}
\affiliation{Institute of Applied Mathematics, Academy of Mathematics and Systems Science, Chinese Academy of Sciences, Beijing 100190, China}
\author{Chun-Yu Lin}
\affiliation{National Center for High-Performance Computing,
Hsinchu 300, Taiwan}
\author{Hsing-Po Pan}
\affiliation{Department of Physics, National Cheng-Kung University,
Tainan 701, Taiwan}
\author{Hwei-Jang Yo}
\affiliation{Department of Physics, National Cheng-Kung University,
Tainan 701, Taiwan}

\date{\today}

\begin{abstract}
Gravitational wave source localization problem is important in gravitational wave astronomy. Regarding ground-based detector, almost all of the previous investigations only considered the difference of arrival time among the detector network for source localization. Within the matched filtering framework, the information beside the arrival time difference can possibly also do some help on source localization. Especially when an eccentric binary is considered, the character involved in the gravitational waveform may improve the source localization. We investigate this effect systematically in the current paper. During the investigation, the enhanced post-circular (EPC) waveform model is used to describe the eccentric binary coalesce. We find that the source localization accuracy does increase along with the eccentricity increases. But such improvement depends on the total mass of the binary. For total mass 100M${}_\odot$ binary, the source localization accuracy may be improved about 2 times in general when the eccentricity increases from 0 to 0.4. For total mass 65M${}_\odot$ binary (GW150914-like binary), the improvement factor is about 1.3 when the eccentricity increases from 0 to 0.4. For total mass 22M${}_\odot$ binary (GW151226-like binary), such improvement is ignorable.
\end{abstract}

\maketitle

\section{\label{intro}introduction}
Along with the gravitational wave detection events GW150914 \cite{PhysRevLett.116.061102}, GW151226 \cite{PhysRevLett.116.241103}, GW170104 \cite{PhysRevLett.118.221101}, and GW170814 \cite{GW170814}, the era of gravitational wave astronomy has come \cite{2017arXiv170300187C}. Coalescing binary compact (CBC) objects are the most promising sources of gravitational waves for the second generation of ground-based interferometric detectors, such as Advanced LIGO (AdvLIGO) \cite{0264-9381-27-8-084006}, Advanced Virgo (AdvVIRGO) \cite{0264-9381-25-18-184001} and KAGRA \cite{0264-9381-29-12-124007}, as well as for the planned third generation detectors like the Einstein Telescope (ET) \cite{0264-9381-26-8-085012} and the Cosmic Explorer (CE) \cite{ClasQuanGra.34.044001}. For example, all GW150914, GW151226, GW170104 and GW170814 are coalescing binary black holes. Although GW150914, GW151226, GW170104 and GW170814 admit vanishing eccentricity orbits, we can expect some gravitational wave signals of eccentric coalescing binary will be detected along with the improvement of the sensitivity of AdvLIGO and other detectors \cite{loutreleccentric}. This is because there are many mechanisms to produce eccentric binary as gravitational wave sources. We just say a few in the following. In globular clusters, stellar mass black hole binaries are expected to have a thermal distribution of eccentricity \cite{Benacquista}. Hierarchical triple systems consist of two closely bound black holes and a third one orbiting the mass center of the first two. This kind of triple system is common in globular clusters. The Kozai mechanism may be significant for these systems. The author in \cite{0004-637X-598-1-419} found that approximately $30\%$ of these binaries could merge with eccentricities $e \gtrsim 0.1$ when they enter the AdvLIGO frequency band. The self-segregate effect of the stellar-mass black holes around a supermassive black hole in galactic nuclei may result in many eccentric BH-BH binaries as AdvLIGO's sources \cite{O'Leary01062009}.

Gravitational wave astronomy is an exciting new window to our Universe. When we use the gravitational wave to observe the objects in the Universe, the accuracy of parsing the parameters of the source is important. Interestingly we found in \cite{LIGOPSD} that many parameters can be got more accurately for an eccentric binary than for a quasi-circular binary. Localizing a gravitational wave source is an important issue in gravitational wave detection. It is a key step in astronomy observation. If ones can not determine the location of the
gravitational wave source accurately enough, ones have to ask other means such as electromagnetic observation to aid. But on the contrary, if ones can localize the gravitational wave source accurately through gravitational wave detection, the multi-messenger astronomy will benefit much \cite{fairhurst2009triangulation}. For space-based detector--LISA, the authors in \cite{PhysRevD.86.104027} found that higher eccentricity does increase the accuracy of source localization. It is interesting to ask how about ground-based detectors. We will investigate this problem in the current paper.

The space-based detectors such as eLISA \cite{amaro2012low}, LISA \cite{audley2017laser}, Taiji \cite{gong2011scientific} and Tianqin \cite{luo2016tianqin}, will change much of their positions within the time scale of the gravitational wave signal. So the information of the gravitational wave source localization will be codded in the output signal of the space-based detector. So the waveform matched filtering can recover the information of the source location. But for ground-based detectors, their positions change very little bit within the gravitational wave signal time scale. In order to localize the gravitational wave sources for LIGO type detectors, people usually use a network of detectors \cite{aasi2013prospects,fairhurst2011source,fairhurst2009triangulation,nissanke2011localizing,PhysRevD.81.082001,chen2015facilitating,PhysRevD.74.082004,Kyutoku01072014}.
Among a detectors network, the signal reaching time is different. People always use this time difference to determine the location of the source. Because the waveform information has nothing to do with the reaching time difference among detectors network, the existing results like the one in \cite{PhysRevD.40.3884} can not use the rich behavior of the waveform for an eccentric binary to improve the source localization.

Recently, there are several waveform models for an eccentric binary have been constructed \cite{PhysRevD.80.084001,PhysRevD.82.024033,PhysRevD.90.084016,PhysRevD.95.024038,Hinderer:2017jcs,PhysRevD.96.044028}. In particular, the enhanced post-circular (EPC) model \cite{tessmerfull,PhysRevD.90.084016} is one of such models. It is interesting to ask if the source location accuracy will increase along with the eccentricity. This is the main topic of the current paper. We will use the enhanced post-circular (EPC) waveform model \cite{PhysRevD.90.084016} to investigate this problem. But as we have tested, the results will keep roughly the same for post-circular waveform model \cite{PhysRevD.80.084001}.

Throughout this paper we will use the geometric units with $c=G=1$. M${}_\odot$ is used to denote the solar mass. The arrangement of this paper is as follows. In Sec.~\ref{sec:EPC}, we introduce EPC waveform model and define variables. Some related information about the network of advanced detectors is also presented there. In Sec.~\ref{sec:FIM}, we describe gravitational wave source localization accuracy estimate method used in the current work. Then in Sec.~\ref{sec:results}, we give the source localization accuracy for eccentric binary. Special attention is paid to the accuracy improvement by the orbit eccentricity. Several factors which affect such improvement including the total mass, position of the source and others are investigated. In the last section, some related discussion is presented. Several complicated formulae are delayed to the appendix.
\section{enhanced post-circular waveform model for ground-based detector network}\label{sec:EPC}
Very recently, the authors in \cite{PhysRevD.80.084001} constructed a waveform model in frequency-domain for an eccentric binary. In that model, the conservative and dissipative orbital dynamics are treated with post-Newtonian approximation. The effect of the small eccentricity is treated through a high-order spectral decomposition. Then the waveform is computed via the method of stationary-phase approximation (SPA). So the authors named the model post-circular (PC) waveform model. Later other authors in \cite{tessmer2010motion,PhysRevD.82.124064,tessmerfull} generalized the above result to higher post-Newtonian (PN) orders. Lately, Huerta \textit{et al.} \cite{PhysRevD.90.084016} extended the post-circular model to enhanced post-circular (EPC) model. This EPC model is designed to reproduce the TaylorF2 model at 3.5 PN order in the zero-eccentricity limit and to reproduce the PC model to leading order in the small eccentricity limit. In \cite{LIGOPSD} we have used EPC model to investigate the parameters estimation for an eccentric binary.

All the waveform expressions for EPC model shown in the literature are for a single detector. Here in order to make our discussion self-contained, we explicitly write out the EPC waveform model for detectors network. We setup an earth coordinate as following. The z-axis is along the spin direction of the earth. The x-axis is along the 0 longitude direction. And the y-axis is determined through right-handed screw rule. EPC model involves 11 parameters, which are $e_{0}, D_{Le}, \mathcal{M}, \eta, t_{ce}, \phi_{c}, \iota_{e}, \beta_{e}, \psi_{e}, \theta_{e}$ and $\phi_{e}$, where $e_{0}$ is the initial eccentricity, $D_{Le}$ is the luminosity distance between the center of the earth and the gravitational-wave source, $\mathcal{M}$ is the chirp mass, $\eta$ is the symmetric mass ratio, $t_{ce}$ is the arrival time of the coalescence signal respect to the center of the earth, $\phi_{c}$ is the orbital phase of coalescence, $\psi_e$ is the polarization angle respect to the earth coordinate described above, $\iota_e$ and $\beta_e$ are the polar angles of the orbital plane respect to the earth coordinate. $\theta_{e}$ and $\phi_{e}$ are the localization spherical angles of the gravitational wave source respect to the center of the earth. And more, we assume the $i$-th detector among the network locates at (altitude, longitude) = $(\theta_i,\phi_i)$, and the arm rotates from north direction to west direction with $\psi_i$. Related to usual notation, altitude $\alpha$N (N means North) corresponds to $\theta_i=\frac{\pi}{2}-\alpha$; altitude $\alpha$S (S means South) corresponds to $\theta_i=\frac{\pi}{2}+\alpha$; longitude $\alpha$E (E means East) corresponds to $\phi_i=\alpha$; and longitude $\alpha$W (W means West) corresponds to $\phi_i=2\pi-\alpha$. The angle $\psi_i$ describes the direction of the x-arm of the detector corresponding to the usual notation N$\psi_i$W. Notation N$\psi_i$W means a direction with angle $\psi_i$ rotating from North to West.

The original EPC waveform involves 11 parameters \cite{PhysRevD.90.084016,LIGOPSD}, including $e_{0}, D_{L}, \mathcal{M}, \eta, t_c, \phi_{c}, \iota, \beta, \psi, \theta$ and $\phi$, where $D_{L}$ is the luminosity distance between the detector and the gravitational-wave source, $t_c$ is the arrival time of the coalescence signal respect to the detector, $\phi_{c}$ is the orbital phase of coalescence, $\psi$ is the polarization angle respect to the detector, $\iota$ and $\beta$ are the polar angles respect to the detector. $\theta$ and $\phi$ are the localization spherical angles of the source respect to the detector. These detector based quantities can be related to earth based quantities mentioned in the last paragraph through
\begin{widetext}
\begin{align}
D_{L}&\approx D_{Le}, t_c\approx t_{ce}, \iota\approx\iota_e, \beta\approx\beta_e, \cos\theta\approx\sin\theta_{e}\sin\theta_{i}\cos(\phi_{e}-\phi_{i})+\cos\theta_{e}\cos\theta_{i},\label{DEr_Da}\\
\tan\phi&=\frac{\sin\theta_{e}\sin\psi_i\cos\theta_{i}\cos(\phi_{e}-\phi_{i})-\cos\theta_{e}\sin\theta_{i}
\sin\psi_i-\sin\theta_{e}\cos\psi_i\sin(\phi_{e}-\phi_{i})}
{-\sin\theta_{e}\cos\psi_i\cos\theta_{i}\cos(\phi_{e}-\phi_{i})+\cos\theta_{e}\sin\theta_{i}\cos\psi_i +\sin\theta_{e}\sin\psi_i\sin(\phi_{i}-\phi_{e})}, \\
\cos\psi&\approx\sin\theta_{e}\sin\psi_{e}\sin\theta_{i}\sin(\phi+\psi_i)-\cos(\phi+\psi_i)\cos\psi_{e}
\cos(\phi_{e}-\phi_{i})-\cos(\phi+\psi_i)\sin\psi_{e}\cos\theta_{e}\sin(\phi_{e}-\phi_{i}) \notag \\
&-\sin(\phi+\psi_i)\cos\psi_{e}\cos\theta_{i}\sin(\phi_{e}-\phi_{i})+\cos\theta_{e}\sin\psi_{e}
\cos\theta_{i}\sin(\phi+\psi_i)\cos(\phi_{e}-\phi_{i}).\label{DEr_psia}
\end{align}
\end{widetext}
These relations result from Eqs. (\ref{DEr_D}), (\ref{DEr_t}), (\ref{DEr_iota}), (\ref{DEr_beta}), (\ref{DEr_theta}), (\ref{DEr_phi}) and (\ref{DEr_psi}). In practice, the gravitational wave sources for ground based detectors locate farther than 100Mpc, so $\frac{R_e}{D_{Le}}<10^{-21}$, where $R_e$ is the radius of the earth. In the above relations we have neglected this small quantity.

Based on the above relation, the EPC waveform model for a detector network with $N$ detectors admits $11$ to-be-determined parameters which include $e_{0}, D_{Le}, \mathcal{M}, \eta, t_{ce}, \phi_{c}, \iota_{e}, \beta_{e}, \psi_{e}, \theta_{e}$, $\phi_{e}$, and $3N$ given parameters $\theta_i, \phi_i, \psi_i$ corresponding to each detector.

\begin{figure}
\centering
\includegraphics[width=0.5\textwidth]{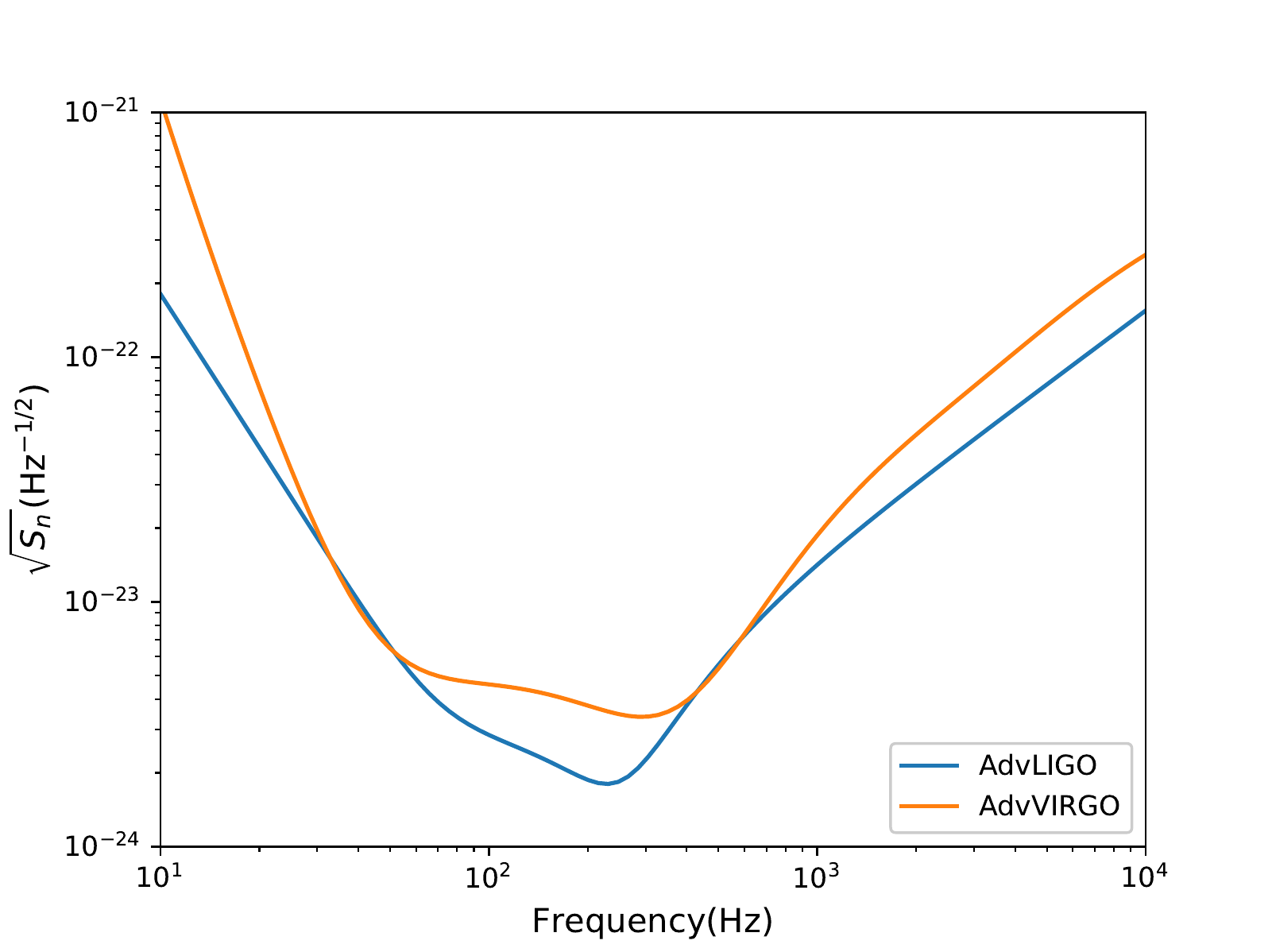}
\caption{The sensitivity curves for Advanced LIGO and Advanced Virgo. The plots correspond to $\sqrt{S_{n}(f)}$ shown in Eqs.~(\ref{snligo}) and (\ref{snvirgo}).}
\label{fig:PSD}
\end{figure}

In the current paper we consider three advanced GW detectors, i.e., two LIGO observatories \cite{LIGO} including the one in Hanford, Washington and the one in Livingston, Louisiana, as well as the VIRGO detector \cite{VIRGO} in Cascina, Italy. The basic information of these three detectors are listed in Table.~\ref{det} for completeness. We approximate the power spectrum density (PSD) of AdvLIGO's sensitivity as \cite{LIGOPSD}
\begin{align}
S_n(f)=S_0\left[x^{-4.14}-5 x^{-2}+\frac{111(1-x^2+x^4/2)}{1+x^2/2}\right]\label{snligo}
\end{align}
when $f>10$Hz, where $x=f/f_0$, $f_0=215$Hz, and $S_0 = 10^{-49}$Hz${}^{-1}$. When $f<10$Hz, $S_n(f)=\infty$. Regarding the PSD for AdvVIRGO's sensitivity, we use approximation \cite{VIRGOPSD}
\begin{equation}
\label{snvirgo}
\begin{split}
S_{n}(f)=&S_0\times[ \\
&0.07\exp(-0.142-1.437x+0.407x^{2}) \\
+&3.10\exp(-0.466-1.043x-0.548x^{2}) \\
+&0.40\exp(-0.304+2.896x-0.293x^{2}) \\
+&0.09\exp(1.466+3.722x-0.984x^{2})]^2,
\end{split}
\end{equation}
when $f>10$Hz, where $x=\ln(f/f_0)$, $f_0=300$Hz, and $S_0 = 1.585081\times10^{-48}$Hz${}^{-1}$. When $f<10$Hz, $S_n(f)=\infty$. The two sensitivity curves $\sqrt{S_{n}(f)}$ for AdvLIGO and AdvVIRGO used in the current work are plotted in Fig.~\ref{fig:PSD}. We can see the most sensitive frequency locates at about 250Hz.
\begin{table}
    \centering
    \caption{The location of detectors and the orientation of their arms \cite{PhysRevD.90.024053,cai2017localization,LIGOVVIRGOinf}. Notation N$36^{\circ}$W means a direction between local North and West, 36 degrees away from North.}
    \begin{tabular}{c c c c c} \hline\hline
        \multirow{2}{*}{Detector} &  \multirow{2}{*}{Latitude}  & \multirow{2}{*}{Longitude} & \multicolumn{2}{c}{Azimuth}  \\ \cline{4-5}
                                            &&          & X arm & Y arm \\ \hline
        LHO(Hanford) & $46^{\circ}27'19"$N & $119^{\circ}24'28"$W & N$36^{\circ}$W & W$36^{\circ}$S \\ \hline

        LLO(Livingston) & $30^{\circ}33'46"$N & $90^{\circ}46'27"$W & W$18^{\circ}$S & S$18^{\circ}$E \\ \hline

        VIRGO & $43^{\circ}37'53"$N & $10^{\circ}30'16"$E & N$19^{\circ}$E & W$19^{\circ}$N \\ \hline\hline

     \end{tabular}
\label{det}
\end{table}

\begin{figure}
\begin{tabular}{c}
\includegraphics[width=0.5\textwidth]{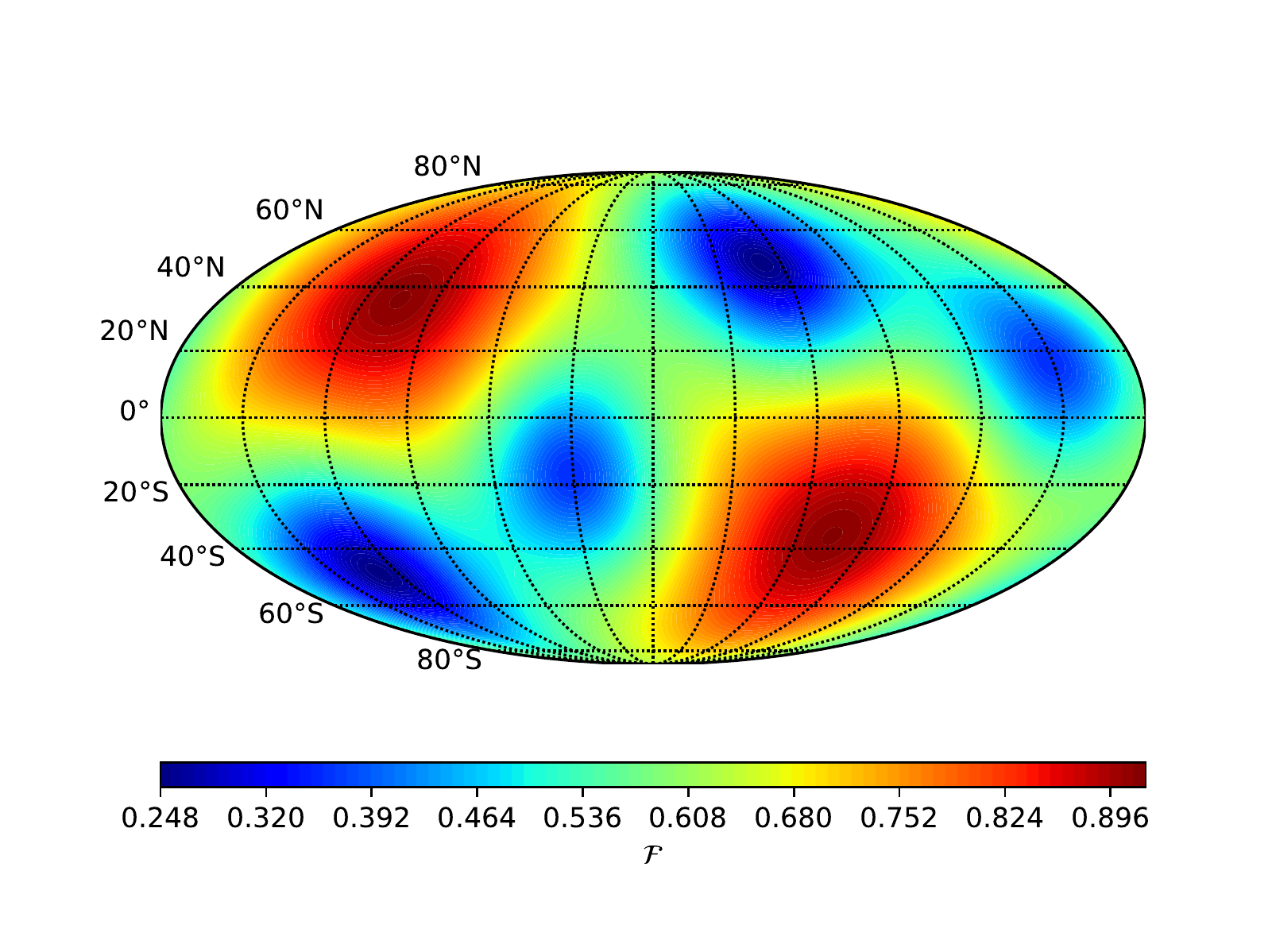}
\end{tabular}
\caption{The antenna pattern function for detectors network composed of the Advanced LIGO in Hanford and in Livingston, and the Advanced VIRGO. The plot corresponds to the pattern function defined in Eq.~(\ref{antenna_w}), where the characteristic frequency $\hat{f}$ is taken as 250Hz.}\label{fig:weighted}
\end{figure}

The antenna beam pattern function $F_{i+}$ and $F_{i\times}$ for each detector, corresponding to gravitational wave's polarization modes $h_+$ and $h_{\times}$, can be expressed as
\begin{align}
&F_{i+}=-\frac{1}{2}(1+\cos^{2}\theta)\cos2\phi\cos2\psi-\cos\theta\sin2\phi\sin2\psi,\notag \\
&F_{i\times}=\frac{1}{2}(1+\cos^{2}\theta)\cos2\phi\sin2\psi-\cos\theta\sin2\phi\cos2\psi.\notag
\end{align}
We relate the quantities $\theta$, $\phi$ and $\psi$ respect to each detector based coordinates to the quantities respect to the earth center based coordinates through Eqs.~(\ref{DEr_D})-(\ref{DEr_psi}) or approximately Eqs.~(\ref{DEr_Da})-(\ref{DEr_psia}). We define the antenna pattern function for the detectors network as
\begin{equation}
\mathcal{F}=\left[\frac{\sum\limits_{i=1}^{N}(F_{i+}^{2}+F_{i\times}^{2})/S_{i}(\hat{f})}{\sum\limits_{i=1}^{N}1/S_{i}
(\hat{f})}\right]^{1/2}\label{antenna_w}
\end{equation}
where $\hat{f}$ is the characteristic frequency for the sensitivity of detector network. For the network composed of Advanced LIGO and Advanced VIRGO, we take $\hat{f}=250$Hz which corresponds to the most sensitive frequency as shown in Fig.~\ref{fig:PSD}. In Fig.~\ref{fig:weighted} we show the antenna pattern function of the detector network we considered.

\section{Gravitational wave source localization accuracy estimate method}\label{sec:FIM}
The measured data $s(t)$ by gravitational wave detector is made up of the signal $h(t)$ and the noise $n(t)$, i.e.,
\begin{equation}
  s(t)=h(t)+n(t).
\end{equation}
Under the assumption that the detector noise is stationary and Gaussian, the likelihood function can be expressed as \cite{PhysRevD.90.062003}
\begin{equation}
   p(s|\theta) \propto e^{-(s-h|s-h)/2},
\end{equation}
where the inner product is defined by
\begin{equation}
  (g|h)=4\text{Re}\int_{0}^{\infty}\frac{\tilde{h}^{*}(f)\tilde{g}(f)}{S_{n}(f)}df.
\end{equation}
Here $\tilde{h}(f)$ and $\tilde{g}(f)$ are the Fourier transforms of $h(t)$ and $g(t)$, ${}^*$ denotes the complex conjugation. And $S_{n}(f)$ is the one-sided power spectral density of the detector noise, which is defined by
\begin{equation}
  \langle n(f)n^{*}(f)\rangle=\frac{1}{2}\delta(f-f^{\prime})S_n(f),
\end{equation}
with $\langle \cdot \rangle$ denotes the probability average respect to the random noise. Considering AdvLIGO and AdvVIRGO's frequency band, we set the lower limit of the above integral as 20Hz. We also assume that the EPC model, as a PN type gravitational wave model, is valid until the last stable orbit frequency, i.e., $\text{F}_{\text{LSO}}\approx\frac{1}{2\pi 6^{3/2}M}$ \cite{PhysRevD.90.084016,LIGOPSD} with $M$ the total mass of the binary. So we use $\text{F}_{\text{LSO}}$ as the upper orbital frequency bound of the integral. For each detector among the network we define
\begin{equation}
  \rho_k^{2}\equiv(h|h)=4\text{Re}\int_{0}^{\infty}\frac{\tilde{h}(f)^{*}\tilde{h}(f)}{S_{kn}(f)}df,
\end{equation}
where $S_{kn}(f)$ is the one-sided power spectral density of the noise for the corresponding $k$th detector. Then the signal to noise ratio (SNR) $\rho_{\text{net}}$ for the network can be expressed as
\begin{equation}
\rho^{2}_{\text{net}}=\sum_{k}\rho^{2}_{k},
\end{equation}
where the summation goes over all of the detectors within the network.

Based on the inner product introduced above, we have the concept Fisher information matrix which is defined as
\begin{equation}
  \bm{\Gamma}_{ij}=\sum_{k}(\partial_{i}h|\partial_{j}h)_{k},
\end{equation}
where the summation again goes over all of the detectors within the network, $\partial_{i}$ means $\partial/\partial p^{i}$ with $p_i$ denotes any parameters among $e_{0}, D_{Le}, \mathcal{M}, \eta, t_{ce}, \phi_{c}, \iota_{e}, \beta_{e}, \psi_{e}, \theta_{e}$ and $\phi_{e}$. So $\bm{\Gamma}_{ij}$ is a 11 by 11 matrix. The Fisher matrix sets a lower bound, i.e., Cramer-Rao lower bound, for the covariance matrix of estimated parameters when the statistical errors are considered, which can be expressed as
\begin{equation}
\text{covar}\left(p_{i},p_{j} \right) \geqslant (\bm{\Gamma}^{-1})_{ij}.
\end{equation}
For a large SNR, the Fisher matrix equals to the covariance matrix. Although in the current paper, we only focus on the location of GW source, i.e., $\theta$ and $\phi$, other parameters should also be considered, because the information on other parameters may also affect the measurement error of $\theta$ and $\phi$. Equivalently, ones can consider Fisher matrix in two-dimensional parameter space $\theta$ and $\phi$ with projection \cite{PhysRevD.81.082001}. With block matrix form
\begin{equation}
\bm{\Gamma}=\left[
\begin{matrix}
\bm{A}&\bm{B}& \\
\bm{B^{T}}&\bm{C}&
\end{matrix}
\right],
\end{equation}
where $\bm{A}$ is the block for $\theta$ and $\phi$. Then the projected Fisher matrix on $\theta$ and $\phi$ space can be expressed as \cite{PhysRevD.81.082001}
\begin{equation}
\bm{\Gamma}_{\text{proj}}=\bm{A}-\bm{B}\bm{C}^{-1}\bm{B^{T}}.
\end{equation}
It can be checked straightforwardly that $(\bm{\Gamma}^{-1}_{\text{proj}})_{ab}=(\bm{\Gamma}^{-1})_{ab}$ where $a$ and $b$ represent $\theta$ and $\phi$.
\begin{figure}
\begin{tabular}{c}
\includegraphics[width=0.5\textwidth]{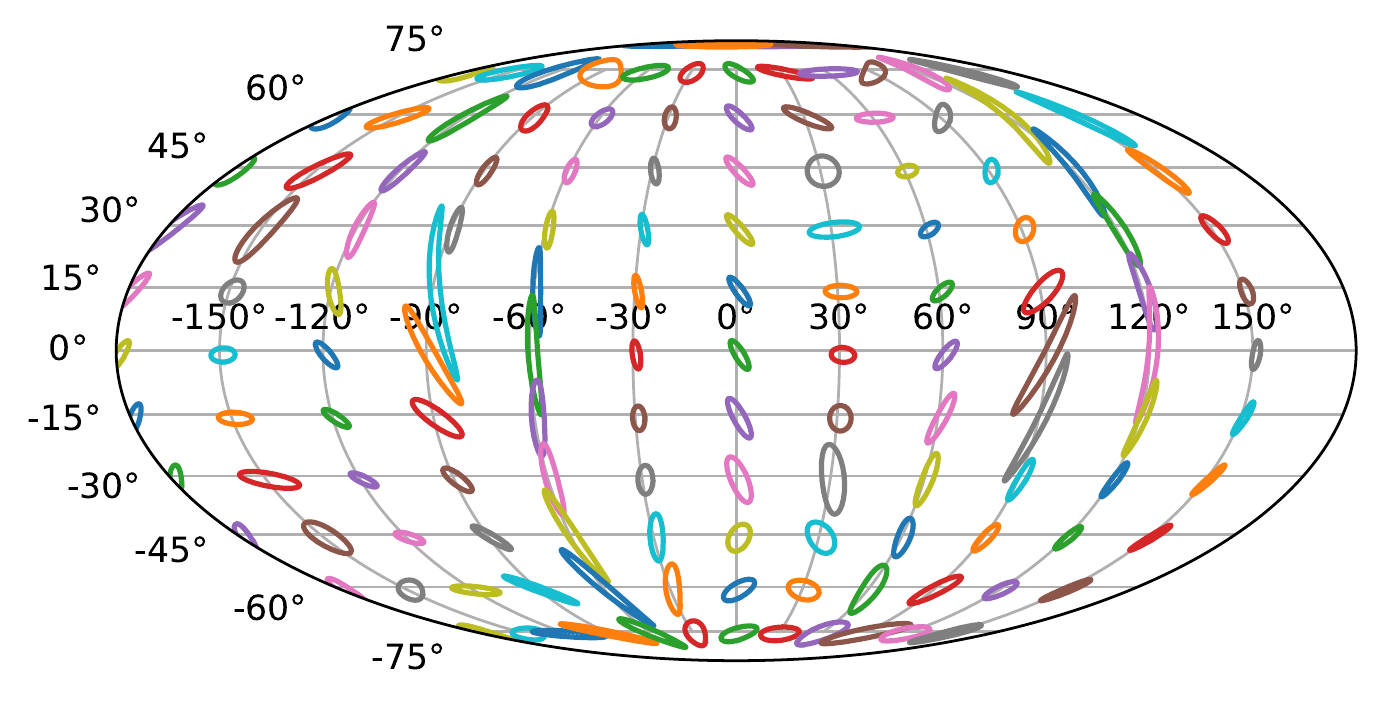}\\
\includegraphics[width=0.5\textwidth]{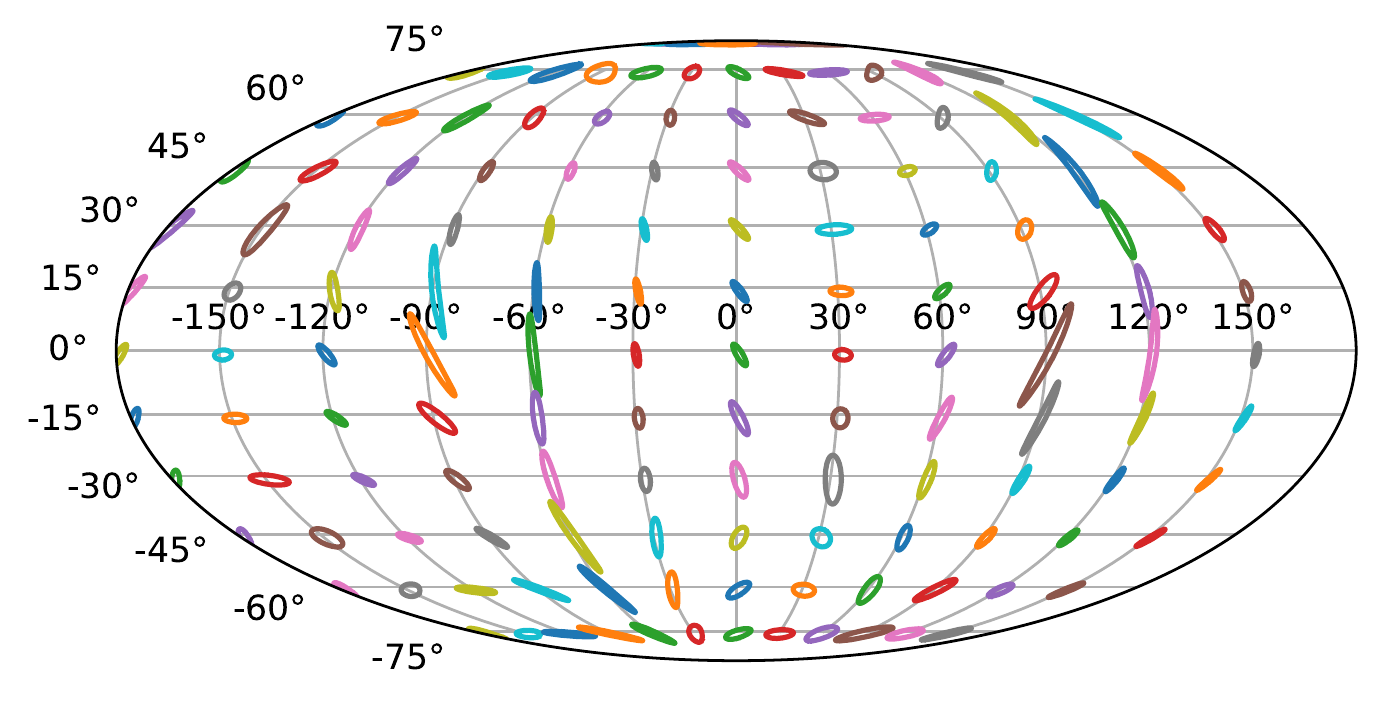}
\end{tabular}
\caption{The source location estimated error ellipse for total mass 100M${}_\odot$ binary black holes. The center of each ellipse represents the corresponding ($\theta_e$, $\phi_e$). The upper and lower plots correspond to $e_0=0$ and $e_0=0.4$ respectively.} \label{fig:100ellipse}
\end{figure}

\begin{figure}
\begin{tabular}{c}
\includegraphics[width=0.5\textwidth]{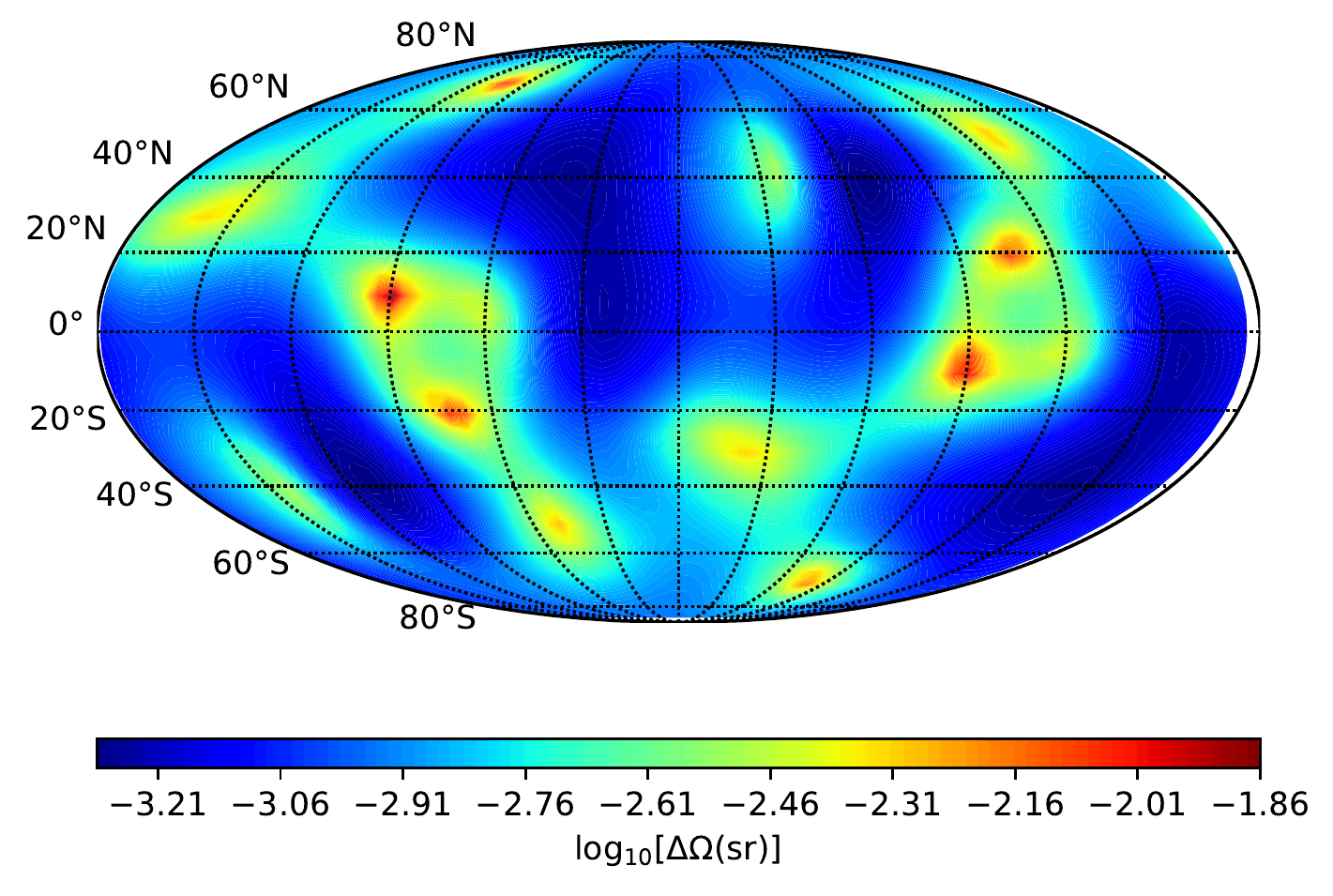}\\
\includegraphics[width=0.5\textwidth]{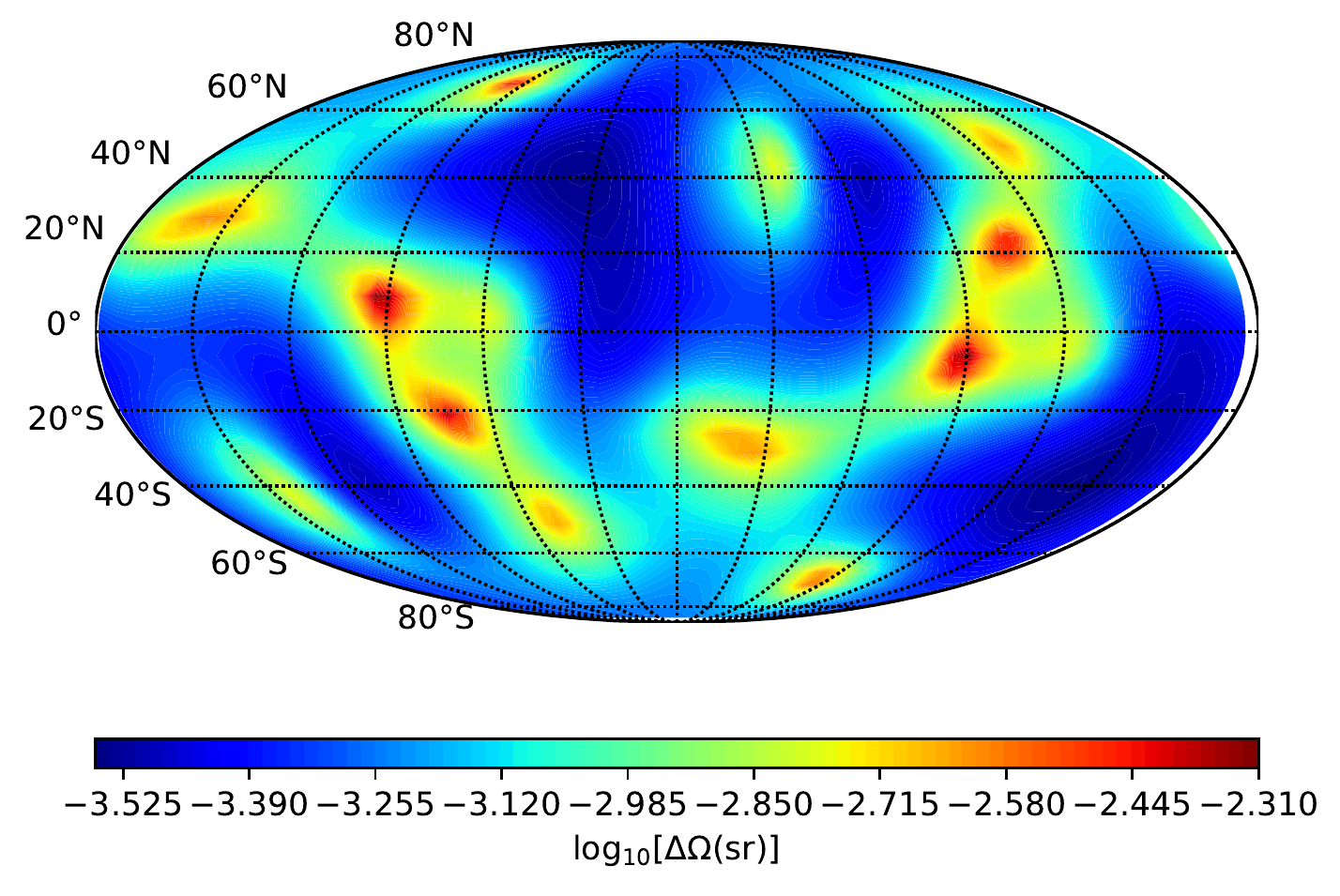}\\
\includegraphics[width=0.5\textwidth]{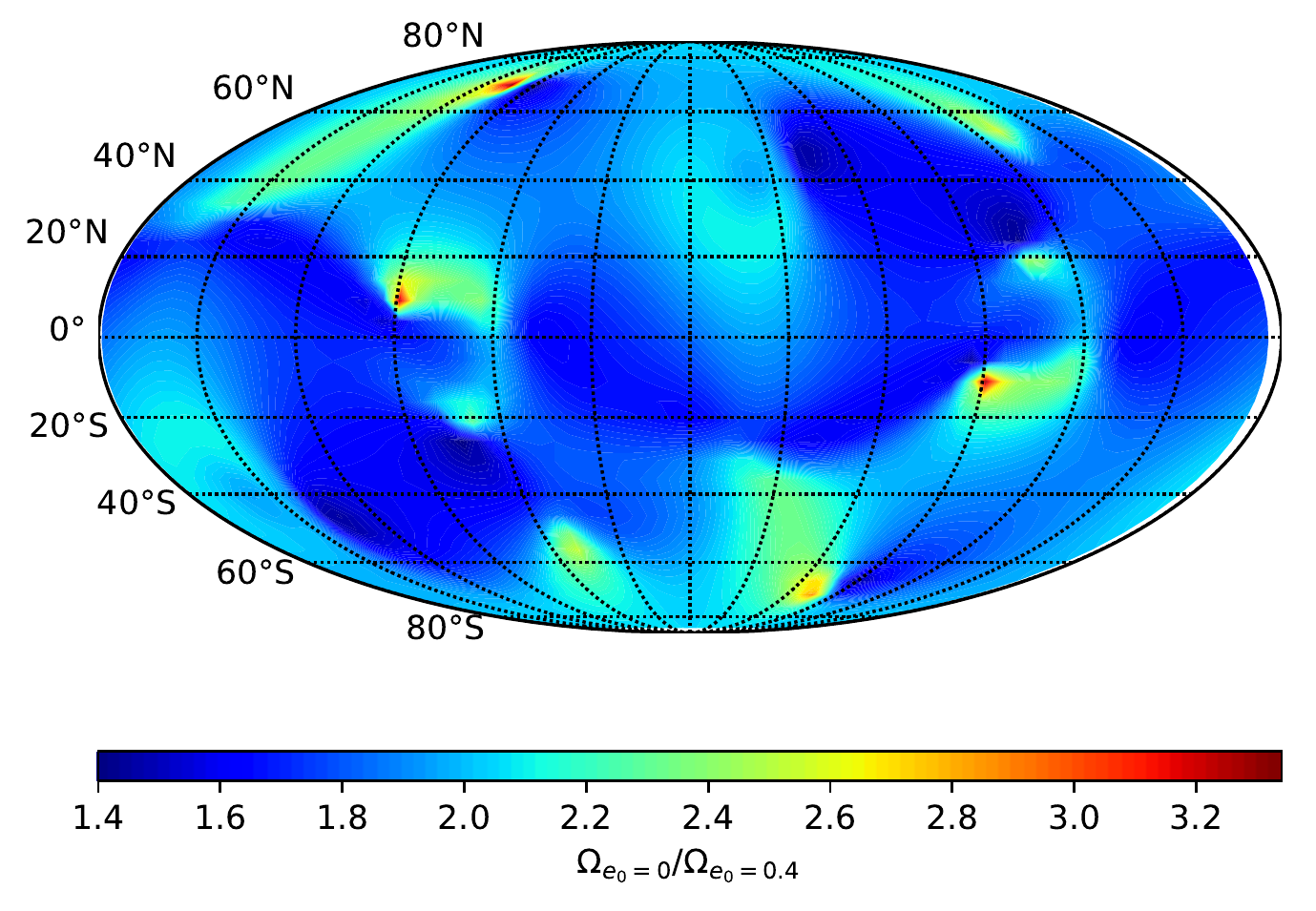}
\end{tabular}
\caption{The source location estimated error $\Delta\Omega$ for binary black hole systems with total mass 100M${}_\odot$. The plots in the upper and the middle panels correspond to $e_0=0$ and $e_0=0.4$ respectively. In the lower plot, we show the relative difference $\frac{\Delta\Omega|_{e_0=0}}{\Delta\Omega|_{e_0=0.4}}$ between $e_0=0$ and $e_0=0.4$. Here sr means square radian, and 1sr $=(180/\pi)^2\approx3282.81$ square degree.} \label{fig:100-space}
\end{figure}

Based on the Fisher matrix we can estimate the parameters measurement error with $\Delta p^{i}=\sqrt{(\bm{\Gamma}^{-1})_{ii}}$. Regarding the source location accuracy, the sky position solid angle is \cite{eccentric}
\begin{equation}
\Delta\Omega=2\pi\sqrt{(\Delta\cos\theta\Delta\phi)^{2}-\langle\Delta\cos\theta\Delta\phi\rangle^{2}}.
\label{eqdomega}
\end{equation}

\section{results for source localization of eccentric binaries}\label{sec:results}
In this section, we use the EPC waveform model described in Sec.~\ref{sec:EPC} to show the influence of eccentricity on the source location accuracy. The EPC waveform model is for a binary black hole (BBH). As indicated by GW150914, GW151226, GW170104 and GW170814, there are possibly many binary black hole systems admit mass range between 10M${}_\odot$ and 100M${}_\odot$. So here we investigate three typical kinds of binary black hole systems with total mass 100M${}_\odot$, 65M${}_\odot$ and 22M${}_\odot$. Respectively we call the 100M${}_\odot$ one the big BBH, 65M${}_\odot$ one GW150914-like BBH and 22M${}_\odot$ one GW151226-like BBH. In this section we consider $e_0$ as the eccentricity of the binary when its orbital frequency is 20Hz.

\begin{figure}
\begin{tabular}{c}
\includegraphics[width=0.5\textwidth]{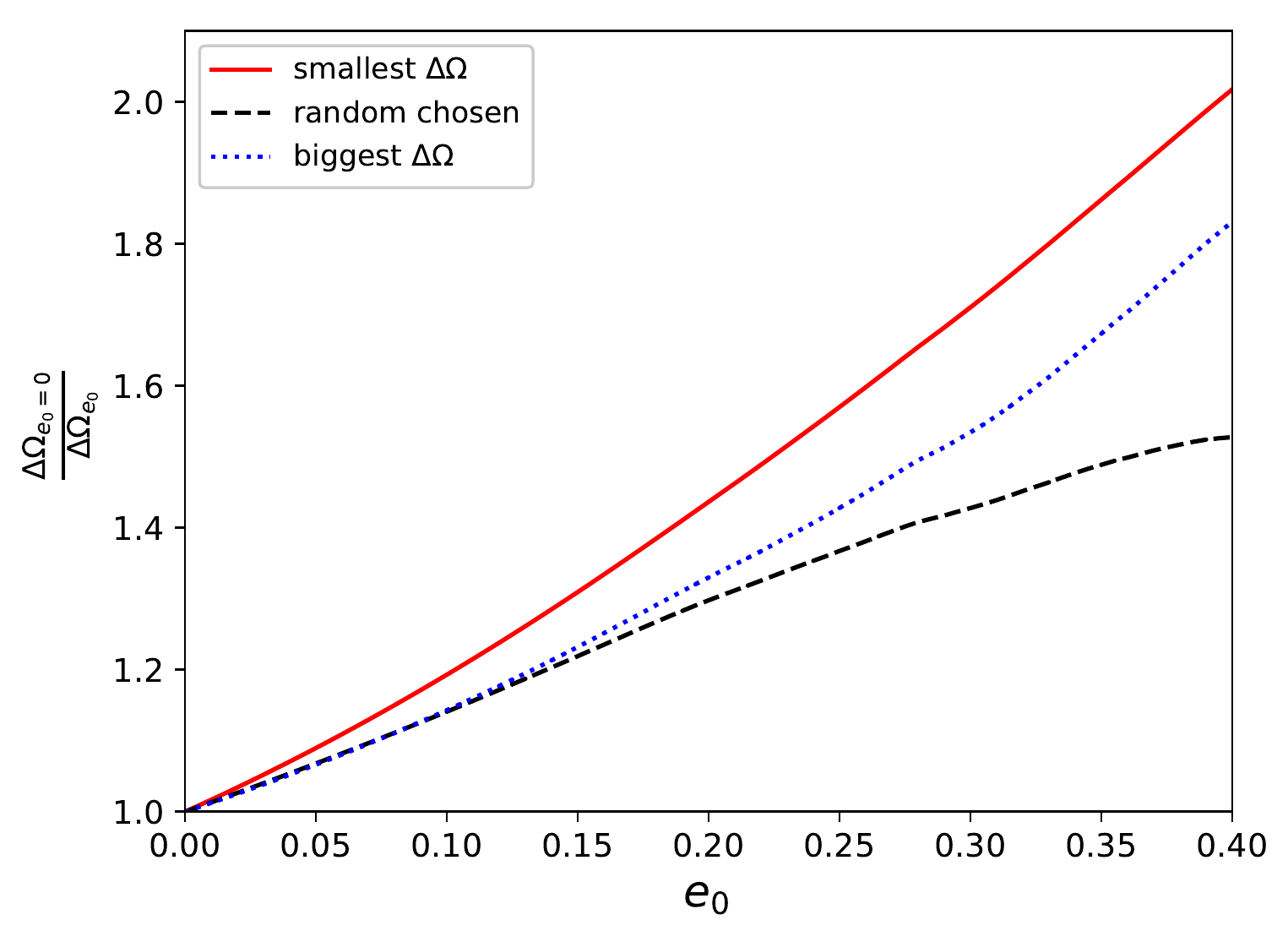}
\end{tabular}
\caption{The improvement factor $\frac{\Delta\Omega|_{e_0=0}}{\Delta\Omega|_{e_0}}$ for big binary black hole systems with total mass 100M${}_\odot$ as the function of initial eccentricity $e_{0}$. The line marked with ``smallest $\Delta\Omega$" corresponds $(\theta_e,\phi_e)=(0.89, 5.69)$ which admits smallest $\Delta\Omega$ in Fig.~\ref{fig:100-space} for $e_0=0.4$. The line marked with ``biggest $\Delta\Omega$" corresponds $(\theta_e,\phi_e)=(1.41, 4.64)$ which admits biggest $\Delta\Omega$ in Fig.~\ref{fig:100-space} for $e_0=0.4$. The line marked with ``random chosen" corresponds $(\theta_e,\phi_e)=(\frac{\pi}{4},\frac{\pi}{4})$ which is chosen arbitrarily. For ``smallest $\Delta\Omega$", ``biggest $\Delta\Omega$" and ``random chosen" cases, $\Delta\Omega|_{e_0=0}$ equals 0.11 square degree, 0.29 square degree, 1.73 square degree respectively.} \label{fig:100-omega-e}
\end{figure}
\subsection{\label{big_BBH}Big BBH case}
Firstly we consider the big binary black hole with total mass 100M${}_\odot$. As an indicative example we fix parameters $D_{Le}=410$Mpc, $\eta=0.25$, $\mathcal{M}=M\eta^{3/5}=43.53$M${}_\odot$, $t_{ce}=0$, $\phi_{c}=0$, $\iota_{e}=0$, $\beta_{e}=0$, $\psi_{e}=0$, while vary $\theta_{e}$, $\phi_{e}$ and $e_0$ to investigate the resulted source location accuracy. In Fig.~\ref{fig:100ellipse}, we compare the results for $e_0=0$ (upper plot) and $e_0=0.4$ (lower plot). Here a series of $\theta_{e}$, $\phi_{e}$ are investigated. The center of each ellipse represents the corresponding ($\theta_e$, $\phi_e$). Each ellipse represents the 3-$\sigma$ error region in the $\theta_{e}$-$\phi_{e}$ parameters space. From Fig.~\ref{fig:100ellipse} we can see that the source location accuracy improves roughly two times when the eccentricity changes from 0 to 0.4.

In Fig.~\ref{fig:100-space} we compare the resulted $\Delta\Omega$ defined in the Eq.~(\ref{eqdomega}) for $e_0=0$ and $e_0=0.4$. Different to Fig.~\ref{fig:100ellipse}, all ($\theta_e$, $\phi_e$) are investigated. The over all distribution behavior of $\Delta\Omega$ respect to ($\theta_e$, $\phi_e$) is similar between $e_0=0$ and $e_0=0.4$ cases. The best source location situation happens at $(\theta_e,\phi_e)=(0.89, 1.22)$ with $\Delta\Omega=5.51\times10^{-4}$sr for $e_0=0$ and at $(\theta_e,\phi_e)=(0.89, 5.69)$ with $\Delta\Omega=2.87\times10^{-4}$sr for $e_0=0.4$. While the worst source location situation happens at $(\theta_e,\phi_e)=(1.41, 4.71)$ with $\Delta\Omega=1.68\times10^{-2}$sr for $e_0=0$ and at $(\theta_e,\phi_e)=(1.41, 4.64)$ with $\Delta\Omega=4.84\times10^{-3}$sr for $e_0=0.4$. In the lower panel of Fig.~\ref{fig:100-space} we plot out the improvement factor $\frac{\Delta\Omega|_{e_0=0}}{\Delta\Omega|_{e_0=0.4}}$ for all ($\theta_e$, $\phi_e$). The $\Delta\Omega$ corresponding to $e_0=0.4$ case improves more than 3 times for the most optimal case compared to that of $e_0=0$, and it improves near 1.5 times for the worst case.

In order to investigate the source location improvement along the increasing of eccentricity $e_0$, we plot the improvement factor $\frac{\Delta\Omega|_{e_0=0}}{\Delta\Omega|_{e_0}}$ respect to $e_0$ in Fig.~\ref{fig:100-omega-e}. Here we consider three situations, one corresponding to the smallest $\Delta\Omega$ in Fig.~\ref{fig:100-space} for $e_0=0.4$ at $(\theta_e,\phi_e)=(0.89, 5.69)$, one corresponding to the biggest $\Delta\Omega$ for $e_0=0.4$ at $(\theta_e,\phi_e)=(1.41, 4.64)$ and the one corresponding to some arbitrarily chosen $(\theta_e,\phi_e)=(\pi/4,\pi/4)$. Based on the results of Figs.~\ref{fig:100-space} and \ref{fig:100-omega-e}, we can infer that the source location may be improved about 2 times in general when the eccentricity increases from 0 to 0.4.

\begin{figure}
\begin{tabular}{c}
\includegraphics[width=0.5\textwidth]{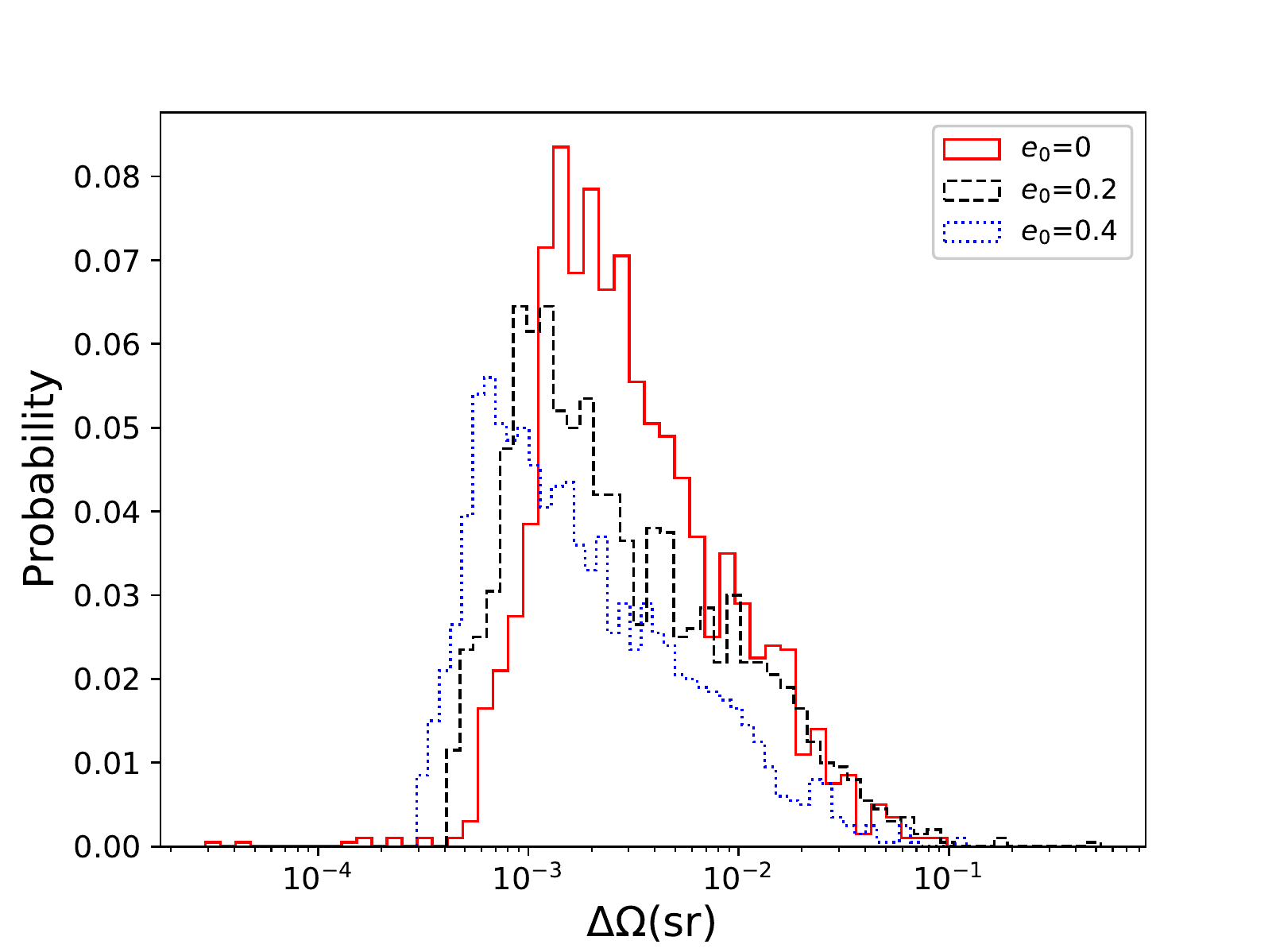}
\end{tabular}
\caption{Histograms of the $\Delta\Omega$ for $10^4$ Monte Carlo sampling of the angle parameters $\iota_{e}, \beta_{e}, \psi_{e}, \theta_{e}$ and $\phi_{e}$. The initial eccentricities $e_0 = 0, 0.2, 0.4$ are considered. This plot is for big binary black hole systems with total mass 100M${}_\odot$.}\label{fig:100-statistic}
\end{figure}
In the above discussions we have fixed $\psi_{e}, \beta_{e}, \iota_{e}$ to 0. In order to investigate the effect of these parameters on the source location improvement by the eccentricity, we investigate $10^4$ samplings through Monte Carlo method. During the sampling process, we take values uniformly for $\theta_e$ in $(0,\pi)$, $\phi_e$ in $(0,2\pi)$, $\psi_e$ in $(0,2\pi)$, $\beta_e$ in $(0,2\pi)$, and $\iota_e$ in $(0,\pi/2)$. The statistics results are shown in Fig.~\ref{fig:100-statistic}.
 While the shape of distribution is insensitive to the eccentricity, as is consistent with other parameter estimation results \cite{LIGOPSD},
 the fact that the peaks of distributions move leftward indicates a better source location accuracy as the eccentricity increases. Compared to $e_0=0$, the source location accuracy of $e_0=0.2$ improves 1.5 times. Compared to $e_0=0.2$, $e_0=0.4$ gets another 1.5 times improvement.

\subsection{GW150914-like BBH case}
\begin{figure}
\begin{tabular}{c}
\includegraphics[width=0.49\textwidth]{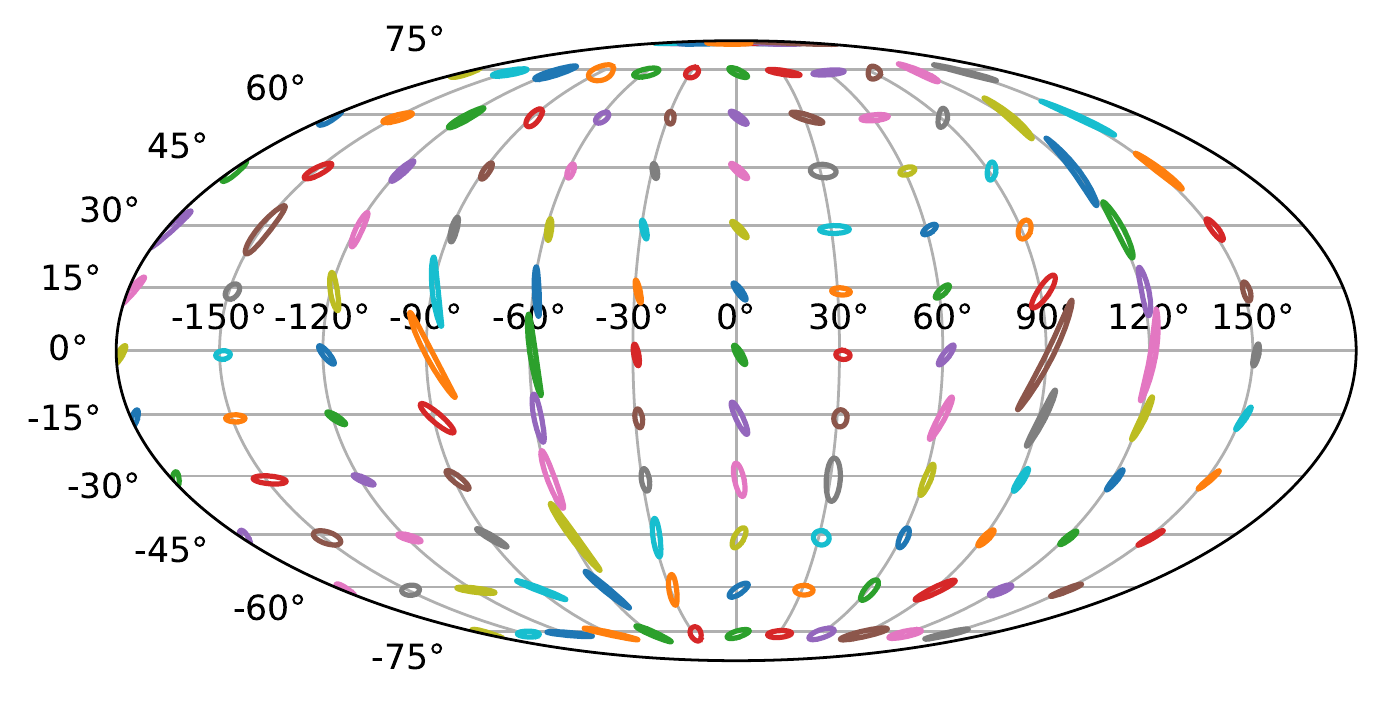}\\
\includegraphics[width=0.49\textwidth]{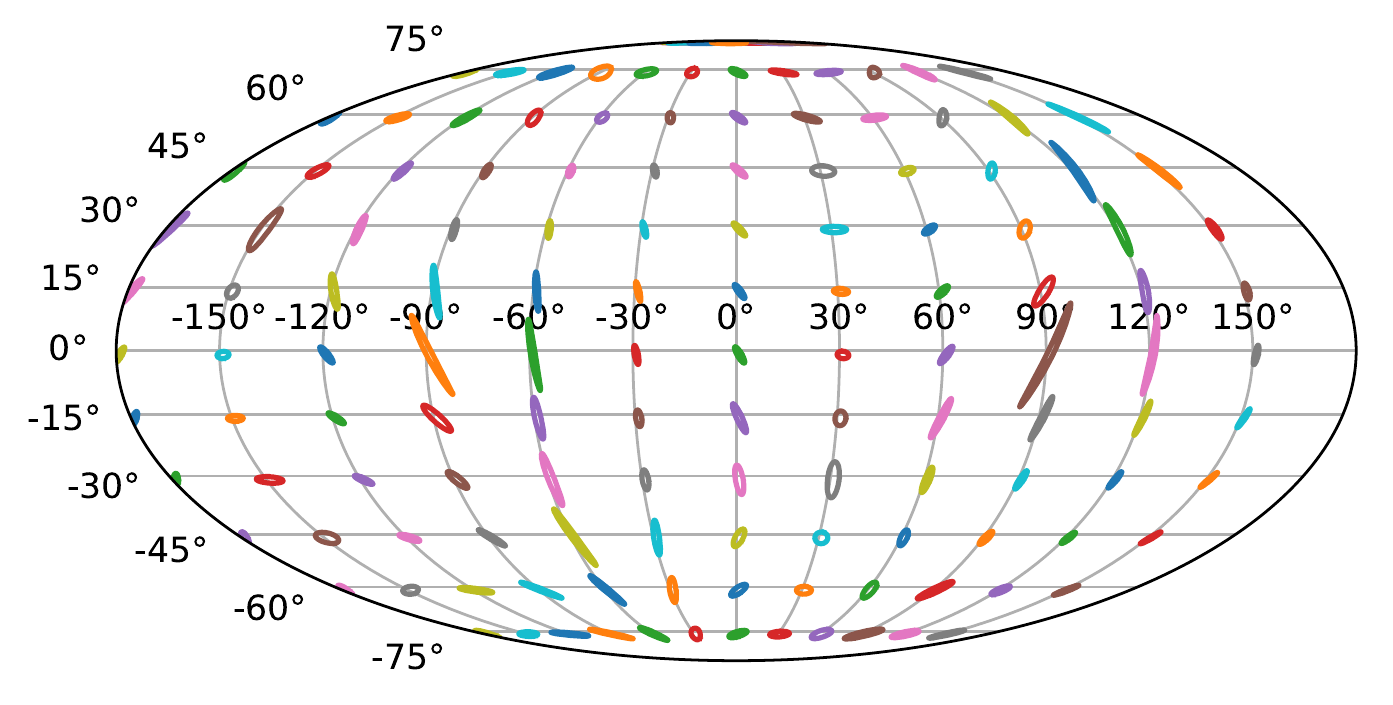}
\end{tabular}
\caption{The source localization error ellipse for the GW150914-like binary black holes with a total mass 65M${}_\odot$.
 The upper and lower plots correspond to $e_0=0$ and $e_0=0.4$, respectively.} \label{fig:065ellipse}
\end{figure}

In this subsection, we consider GW150914-like BBH sources. We set the involved parameters as $D_{Le}=410$Mpc, $\eta=0.25$, $\mathcal{M}=28.3$M${}_\odot$, $t_{ce}=0$, $\phi_{c}=0$, $\iota_{e}=0$, $\beta_{e}=0$, $\psi_{e}=0$. Compared to the setting in the above subsection the only different parameter is the chirp mass $\mathcal{M}$. The location parameters $\theta_{e}$, $\phi_{e}$ and the eccentricity $e_0$ are investigated for different values. Similar to the Fig.~\ref{fig:100ellipse}, we investigate the error ellipses for different $(\theta_e,\phi_e)$. Again the error ellipses represent the 3-$\sigma$ error region in the parameter space. The results are plotted in Fig.~\ref{fig:065ellipse}. Compared to the ellipses in the Fig.~\ref{fig:100ellipse}, we can find that the source localization accuracy is better here. We attribute this to that more signal falls in the LIGO frequency band for GW150914-like binary than for the big BBH.

\begin{figure}
\begin{tabular}{c}
\includegraphics[width=0.47\textwidth]{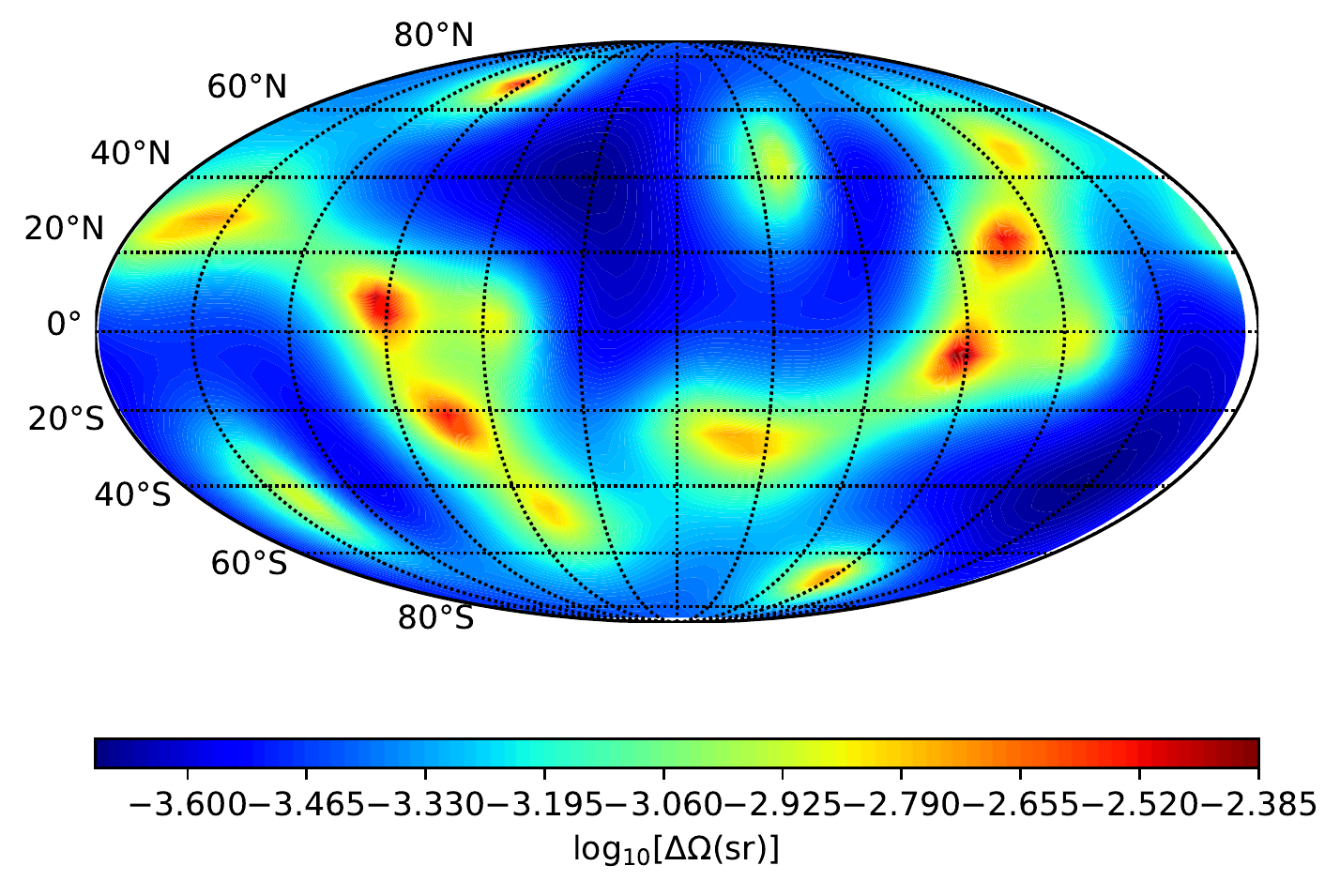}\\
\includegraphics[width=0.47\textwidth]{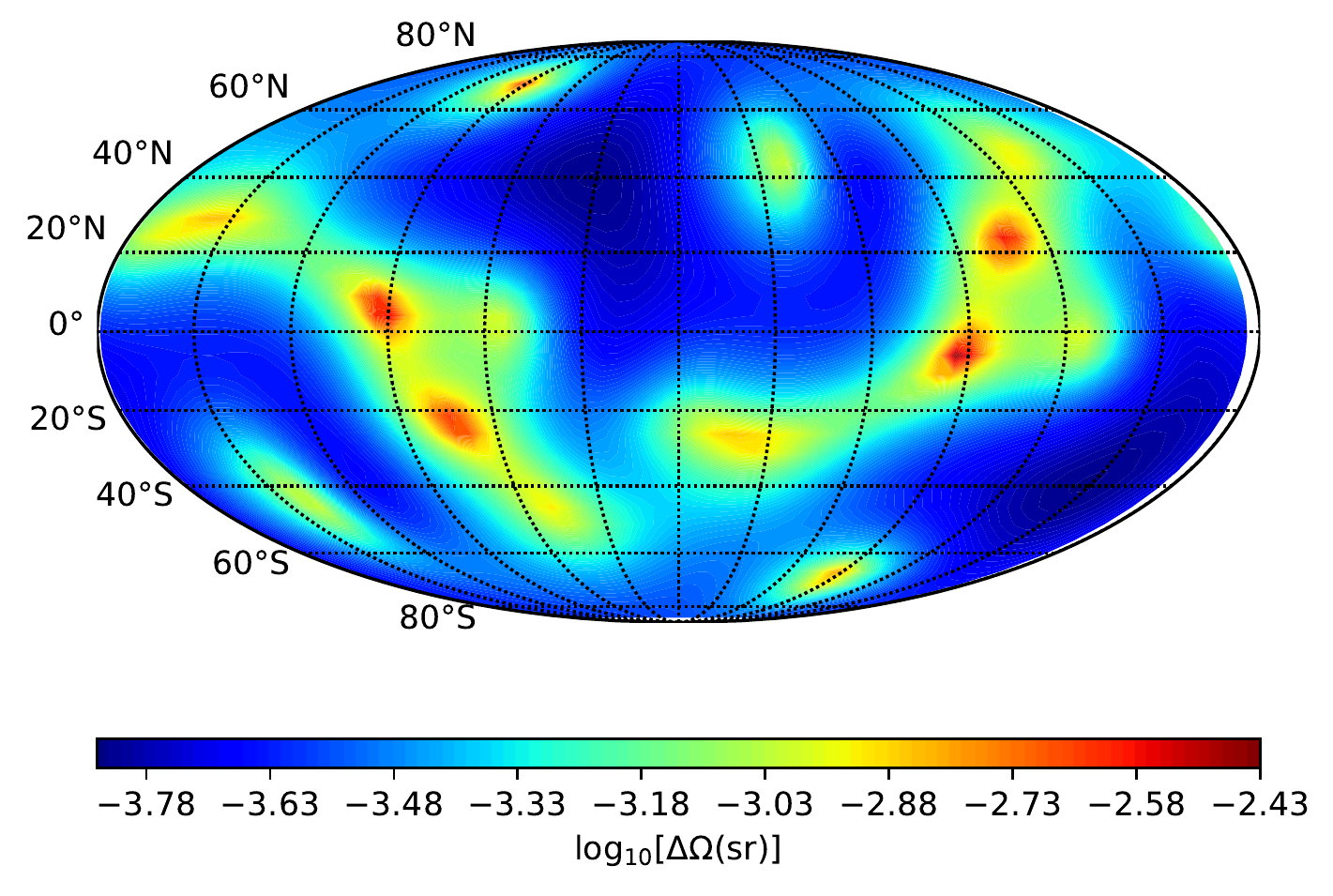}\\
\includegraphics[width=0.47\textwidth]{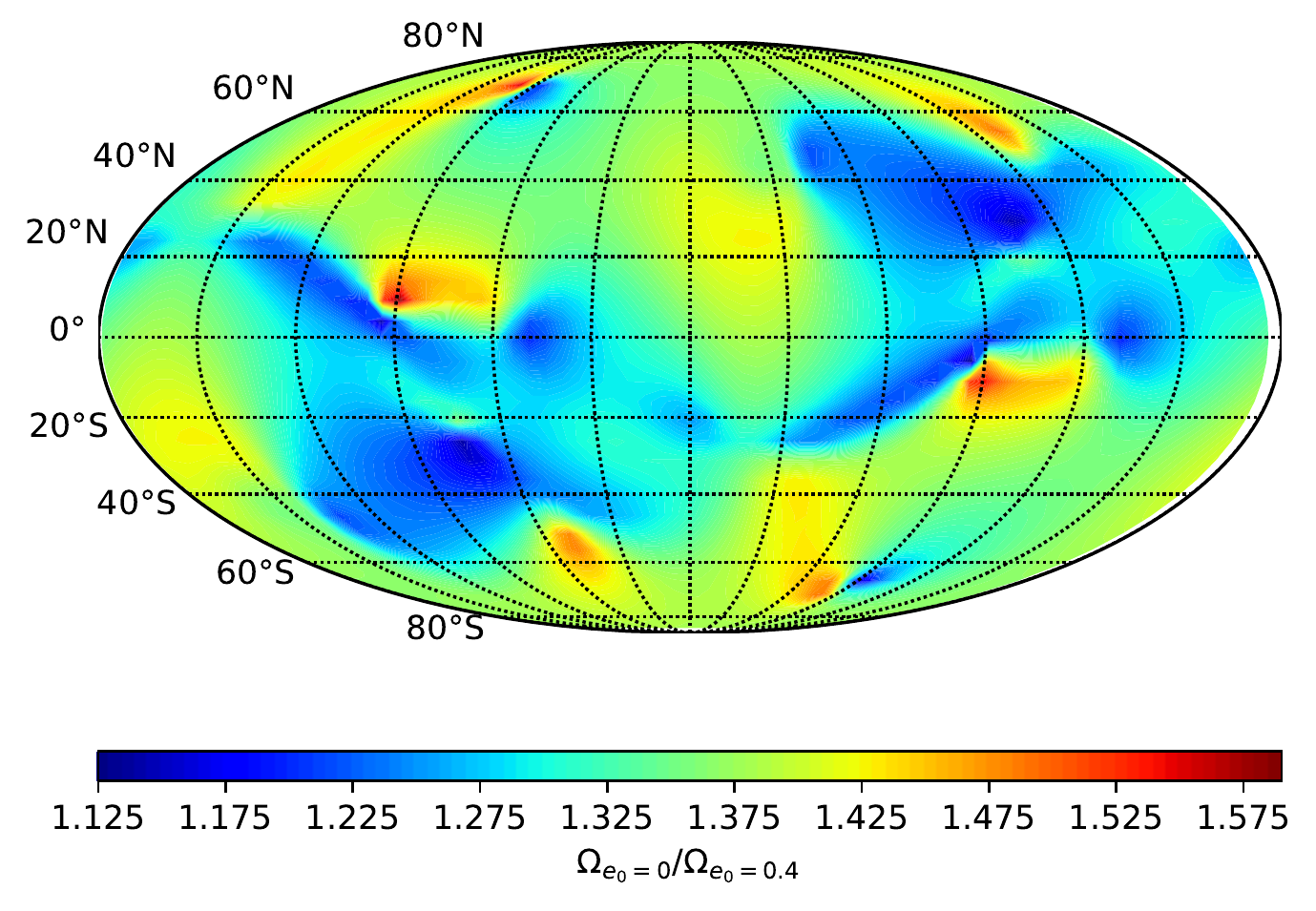}
\end{tabular}
\caption{The source location estimated error $\Delta\Omega$ for GW150914-like binary black holes with total mass 65M${}_\odot$. In the upper and the middle plots $e_0=0$ and $e_0=0.4$ respectively. In the lower plot, we show the relative difference $\frac{\Delta\Omega|_{e_0=0}}{\Delta\Omega|_{e_0=0.4}}$ between $e_0=0$ and $e_0=0.4$.} \label{fig:65-space}
\end{figure}

Similar to the Fig.~\ref{fig:100-space}, we plot the distribution of $\Delta\Omega$ for GW150914-like BBHs in Fig.~\ref{fig:65-space}. Compared to the Fig.~\ref{fig:100-space}, we can find that the distribution behavior is independent of the total mass of BBH and the eccentricity. At the same time, we can also note that the distribution is different to the pattern function shown in Sec.~\ref{sec:EPC}. For the $e_0=0$ and $e_0=0.4$ cases respectively, the source location accuracy is better than that of big BBH cases. This is natural, because more gravitational wave signal falls in the LIGO frequency band compared to the big BBH cases as mentioned above. The best source location case happens at $(\theta_e,\phi_e)=(0.89,5.69)$ with $\Delta\Omega=2.13\times10^{-4}$sr for $e_0=0$ and at $(\theta_e,\phi_e)=(0.89, 5.69)$ with $\Delta\Omega=1.49\times10^{-4}$sr for $e_0=0.4$. The worst source location situation happens at $(\theta_e,\phi_e)=(1.68, 1.50)$ with $\Delta\Omega=4.26\times10^{-3}$sr for $e_0=0$ and at $(\theta_e,\phi_e)=(1.68, 1.50)$ with $\Delta\Omega=3.59\times10^{-3}$sr for $e_0=0.4$. The distribution of $\frac{\Delta\Omega|_{e_0=0}}{\Delta\Omega|_{e_0=0.4}}$ respect to $(\theta_e,\phi_e)$ is also similar to that of Fig.~\ref{fig:65-space}. But the range is among $(1.125,1.575)$. And the $\Delta\Omega$ improves about 1.5 times for the most optimal case when $e_0$ changes from 0 to 0.4, which is smaller than the big BBHs shown in the above subsection.

\begin{figure}
\begin{tabular}{c}
\includegraphics[width=0.47\textwidth]{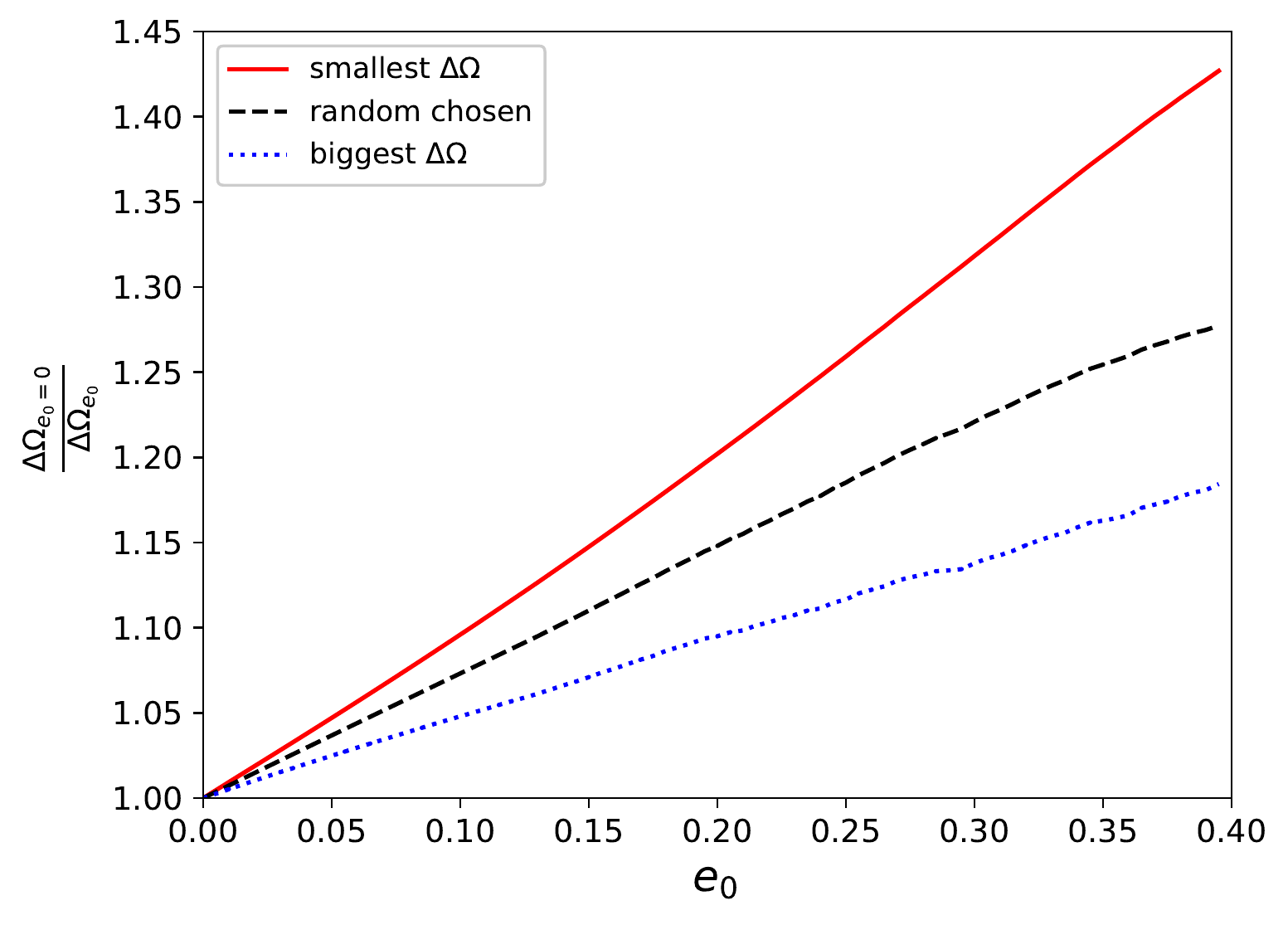}
\end{tabular}
\caption{The improvement factor $\frac{\Delta\Omega|_{e_0=0}}{\Delta\Omega|_{e_0}}$ for GW150914-like binary black holes with total mass 65M${}_\odot$ as the function of initial eccentricity $e_{0}$. The line marked with ``smallest $\Delta\Omega$" corresponds $(\theta_e,\phi_e)=(0.89, 5.69)$ which admits smallest $\Delta\Omega$ in Fig.~\ref{fig:65-space} for $e_0=0.4$. The line marked with ``biggest $\Delta\Omega$" corresponds $(\theta_e,\phi_e)=(1.68, 1.50)$ which admits biggest $\Delta\Omega$ in Fig.~\ref{fig:65-space} for $e_0=0.4$. The line marked with ``random chosen" corresponds $(\theta_e,\phi_e)=(\frac{\pi}{4},\frac{\pi}{4})$ which is chosen arbitrarily. For ``smallest $\Delta\Omega$", ``biggest $\Delta\Omega$" and ``random chosen" cases, $\Delta\Omega|_{e_0=0}$ equals 0.70 square degree, 13.98 square degree, 2.93 square degree respectively.} \label{fig:65-omega-e}
\end{figure}

Fig.~\ref{fig:65-omega-e} gives the improvement factor respect to $e_0$. We again consider three situations, i.e., the smallest $\Delta\Omega$ in Fig.~\ref{fig:65-space} for $e_0=0.4$ at $(\theta_e,\phi_e)=(0.89,5.69)$, the biggest $\Delta\Omega$ in Fig.~\ref{fig:65-space} for $e_0=0.4$ at $(\theta_e,\phi_e)=(1.68,1.50)$ and one arbitrary case at $(\theta_e,\phi_e)=(\pi/4,\pi/4)$. Compare this to the Fig.~\ref{fig:100-omega-e}, we can see that $e_0$ has less influence on source localization error $\Delta\Omega$ for smaller total mass BBHs.

\begin{figure}
\begin{tabular}{c}
\includegraphics[width=0.49\textwidth]{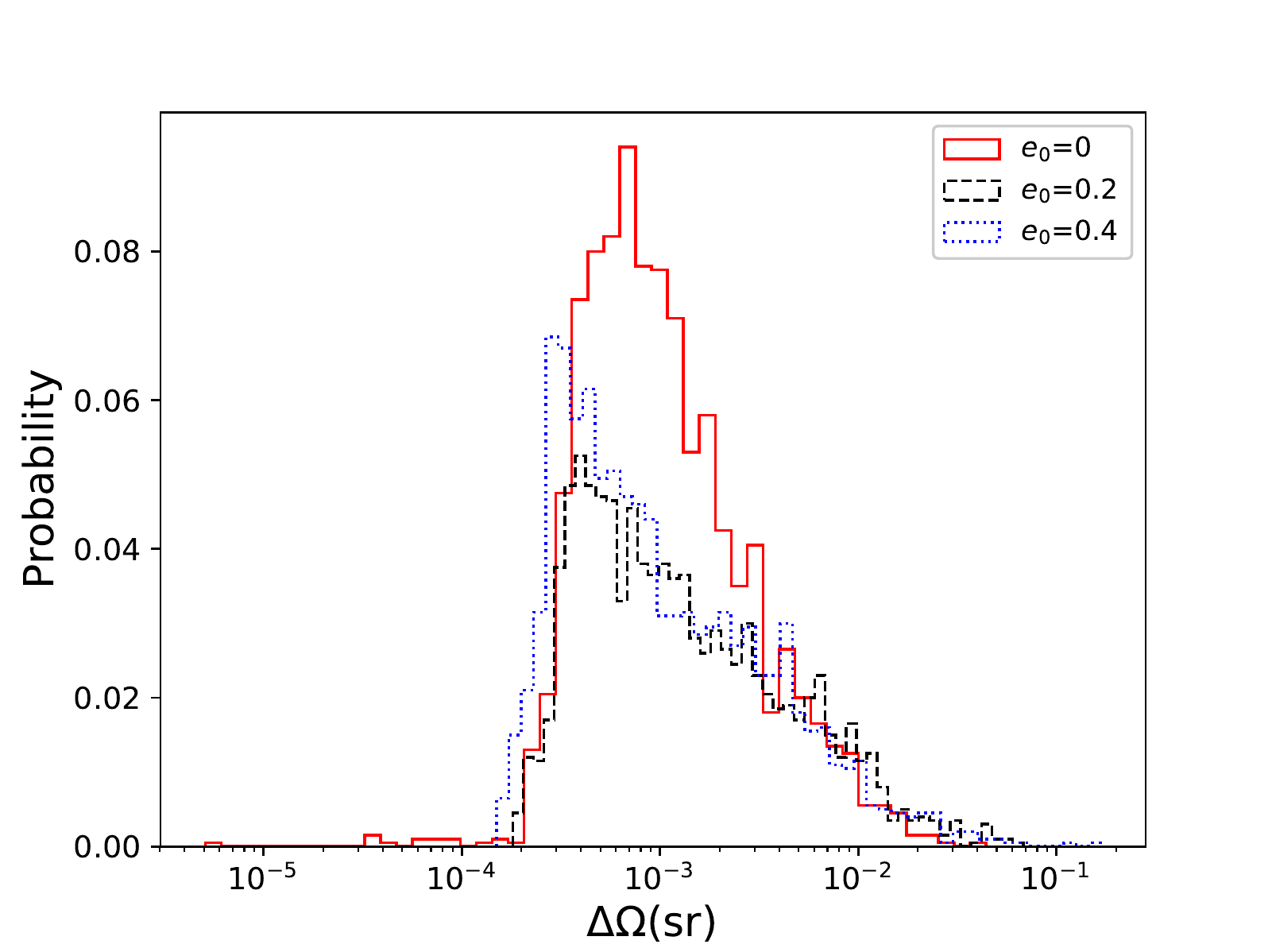}
\end{tabular}
\caption{Histograms of the $\Delta\Omega$ for $10^4$ Monte Carlo sampling of the angle parameters $\iota_{e}, \beta_{e}, \psi_{e}, \theta_{e}$ and $\phi_{e}$. The initial eccentricities $e_0 = 0, 0.2, 0.4$ are considered. This plot is for GW150914-like binary black holes with total mass 65M${}_\odot$.}\label{fig:65-statistic}
\end{figure}

Using Monte Carlo samplings, Fig.~\ref{fig:65-statistic} shows the statistic of $\Delta\Omega$. Compared to $e_0=0$, the source location accuracy of $e_0=0.2$ improves 1.1 times. Compared to $e_0=0.2$, $e_0=0.4$ gets another 1.2 times improvement. This result is consistent to that of the Fig.~\ref{fig:65-omega-e}, and confirms that $e_0$ has less influence on source localization error $\Delta\Omega$ for smaller total mass BBHs.

\subsection{GW151226-like BBH case}
\begin{figure}
\begin{tabular}{c}
\includegraphics[width=0.47\textwidth]{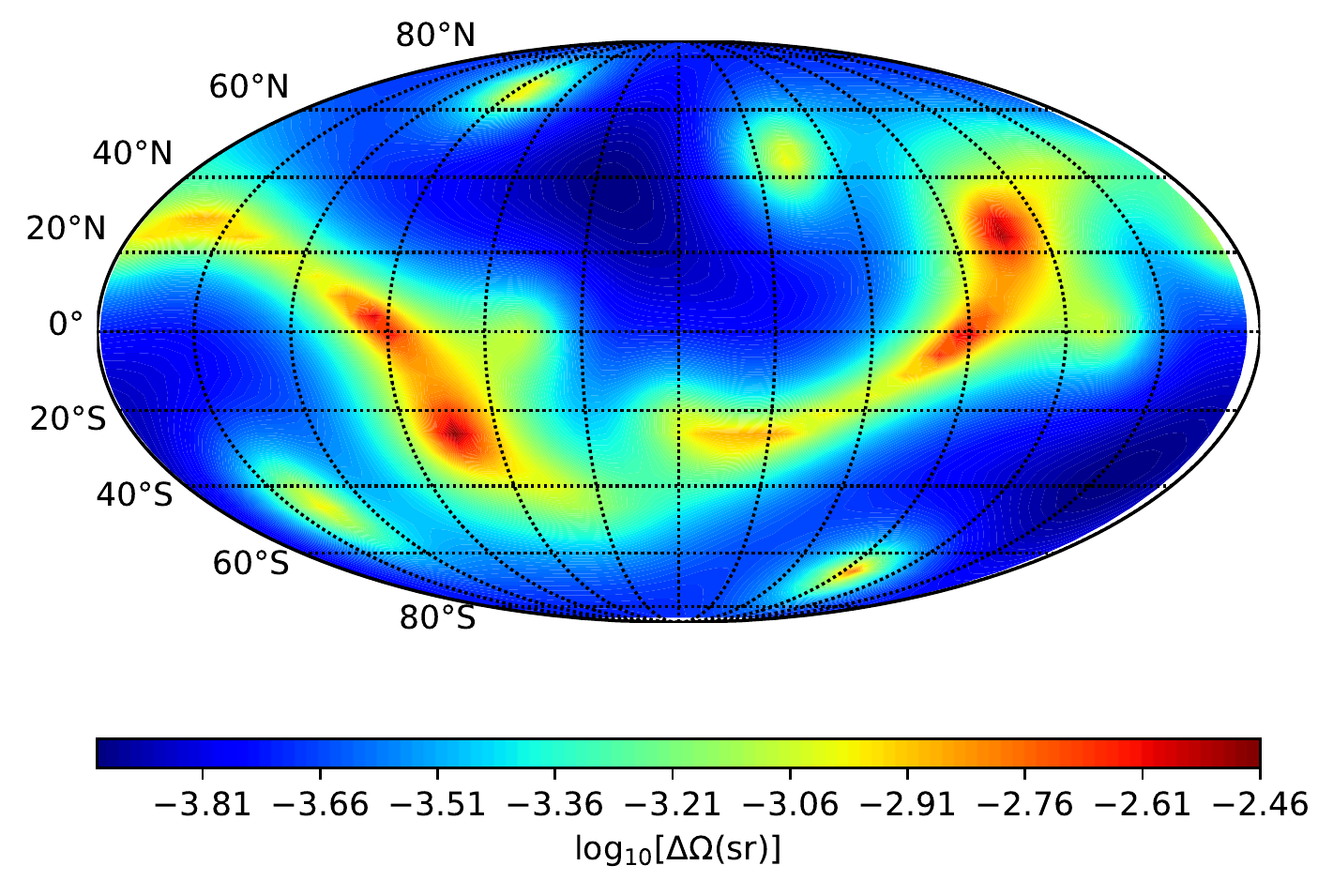}\\
\includegraphics[width=0.47\textwidth]{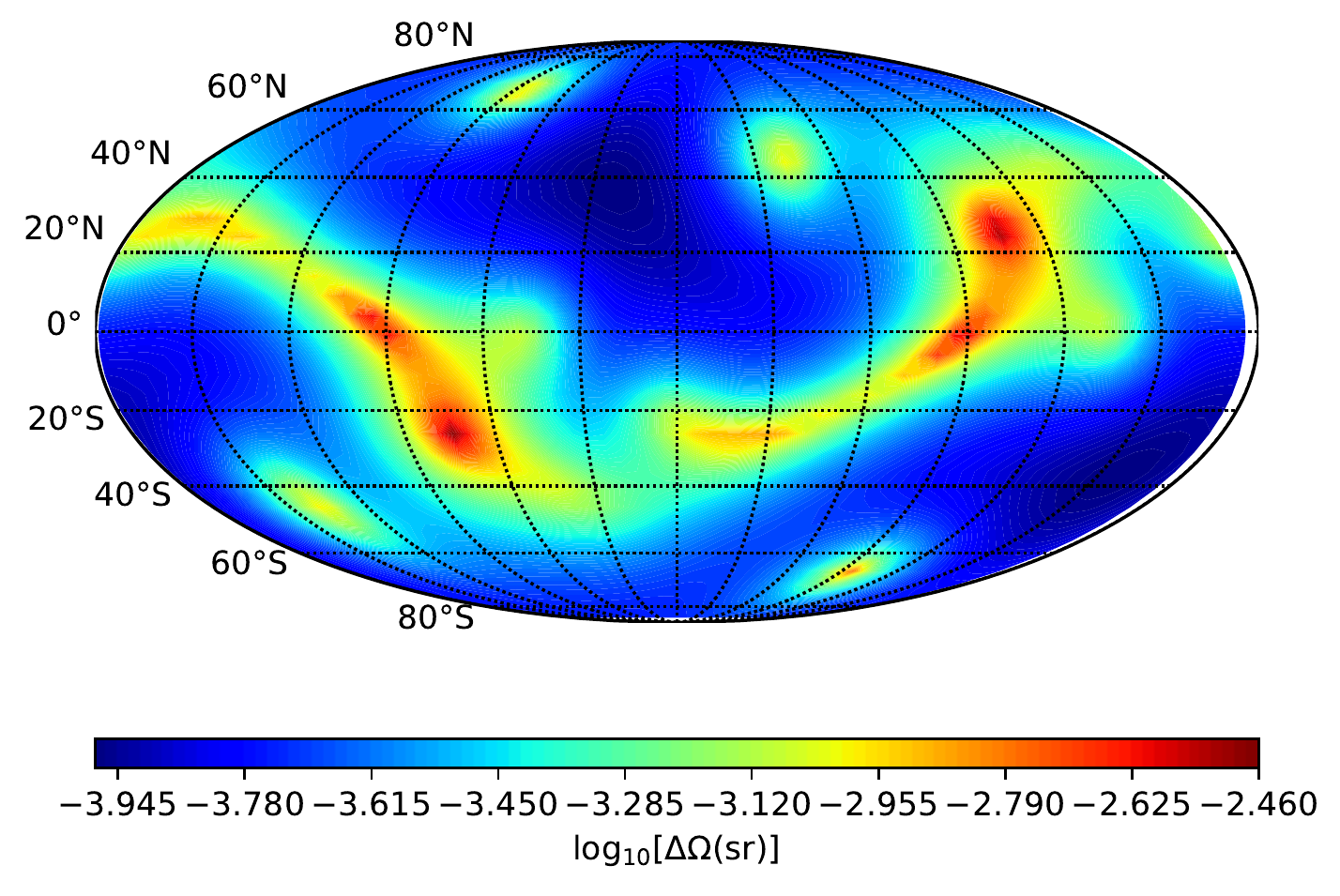}\\
\includegraphics[width=0.47\textwidth]{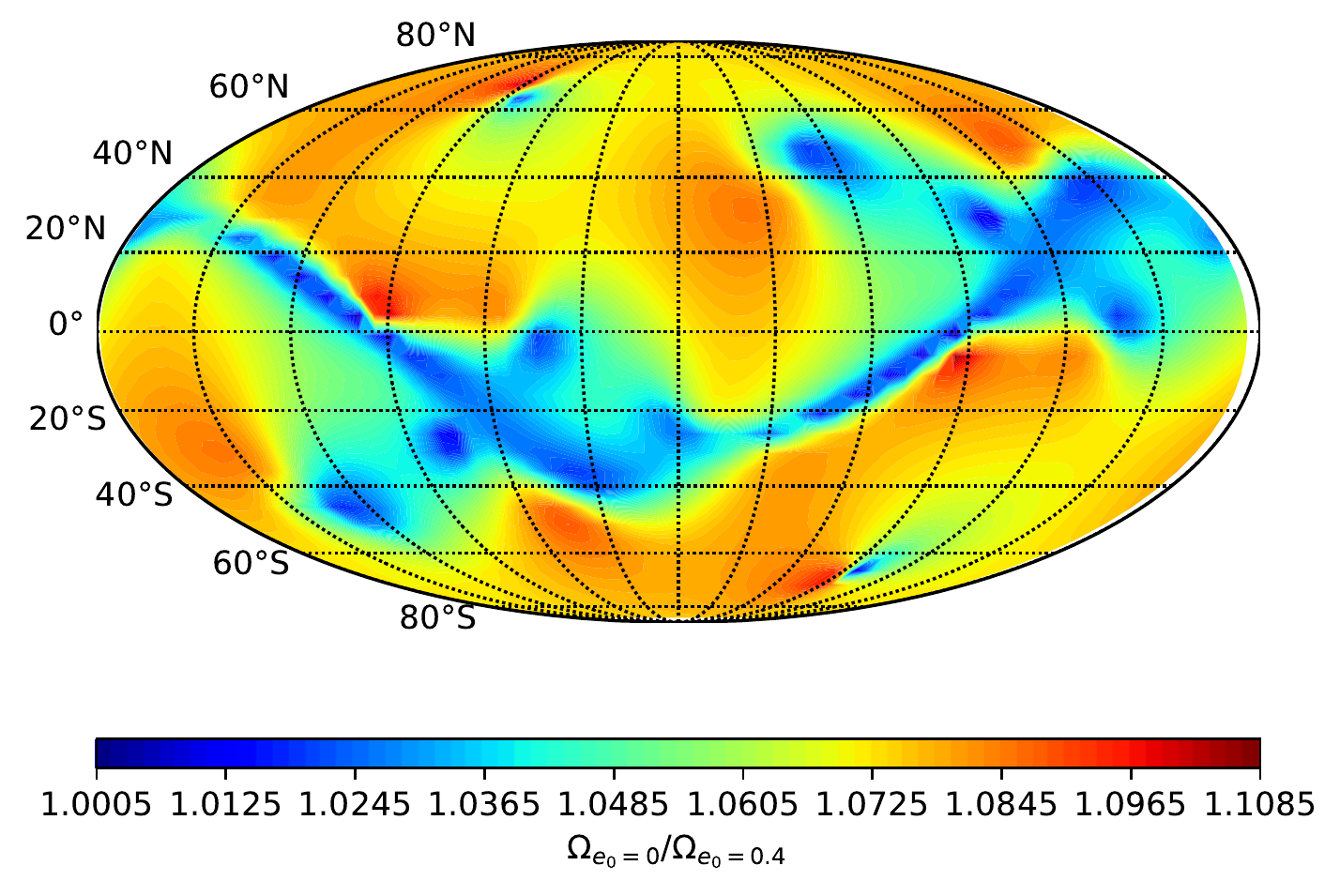}
\end{tabular}
\caption{The source location estimated error $\Delta\Omega$ for GW151226-like binary black holes. In the upper and the middle plots $e_0=0$ and $e_0=0.4$ respectively. In the lower plot, we show the relative difference $\frac{\Delta\Omega|_{e_0=0}}{\Delta\Omega|_{e_0=0.4}}$ between $e_0=0$ and $e_0=0.4$.} \label{fig:22-space}
\end{figure}

In this subsection, we consider GW151226-like BBH sources with total mass 22M${}_\odot$. We set the involved parameters as $D_{Le}=410$Mpc, $\eta=0.25$, $\mathcal{M}=9.58$M${}_\odot$, $t_{ce}=0$, $\phi_{c}=0$, $\iota_{e}=0$, $\beta_{e}=0$, $\psi_{e}=0$. Compared to the setting in the above two subsections the only different parameter is the chirp mass $\mathcal{M}$. Like the GW150914-like BBH case, the location parameters $\theta_{e}$, $\phi_{e}$ and the eccentricity $e_0$ are investigated for different values. For this kind of binary black hole case the improvement on the localization accuracy by the eccentricity is ignorable. Quantitatively we show the distribution behavior of $\Delta\Omega$ for $e_0=0$ and $e_0=0.4$ in the Fig.~\ref{fig:22-space}. The distribution of $\frac{\Delta\Omega|_{e_0=0}}{\Delta\Omega|_{e_0=0.4}}$ respect to $(\theta_e,\phi_e)$ is similar to that of Fig.~\ref{fig:65-space}. But the range is among $(1.0005,1.1085)$. In general, the improvement factor is only 1.05. When the eccentricity is smaller than 0.4, the general improvement factor is even less than 1.05.
\section{DISCUSSION}
Based on enhanced post-circular waveform (EPC) model we have investigated the effect of eccentricity on the source localization accuracy. Along with the analysis process, the detailed matched filtering technique and Fisher information matrix method are adopted. In general, the eccentricity can improve the source localization accuracy. At the same time, we found that the improvement depends on the total mass of the binary. When the total mass is about 100M${}_\odot$, the improvement factor is about 2 times. The improvement factor will decay along with the deceasing of the total mass. When the total mass is about 22M${}_\odot$, the improvement factor decay to about 1.05 which is ignorable.
\begin{figure}
\begin{tabular}{c}
\includegraphics[width=0.5\textwidth]{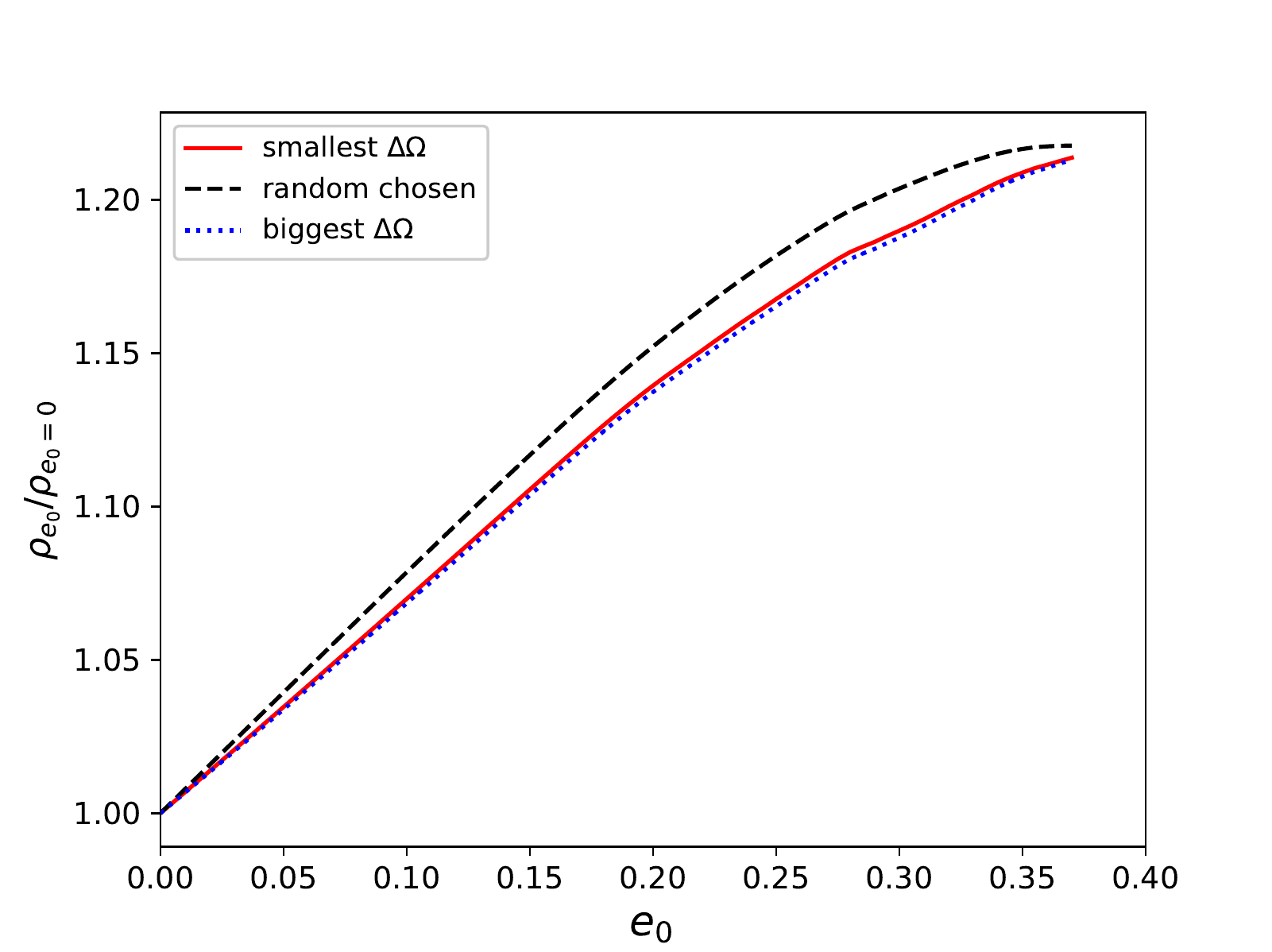}\\
\includegraphics[width=0.5\textwidth]{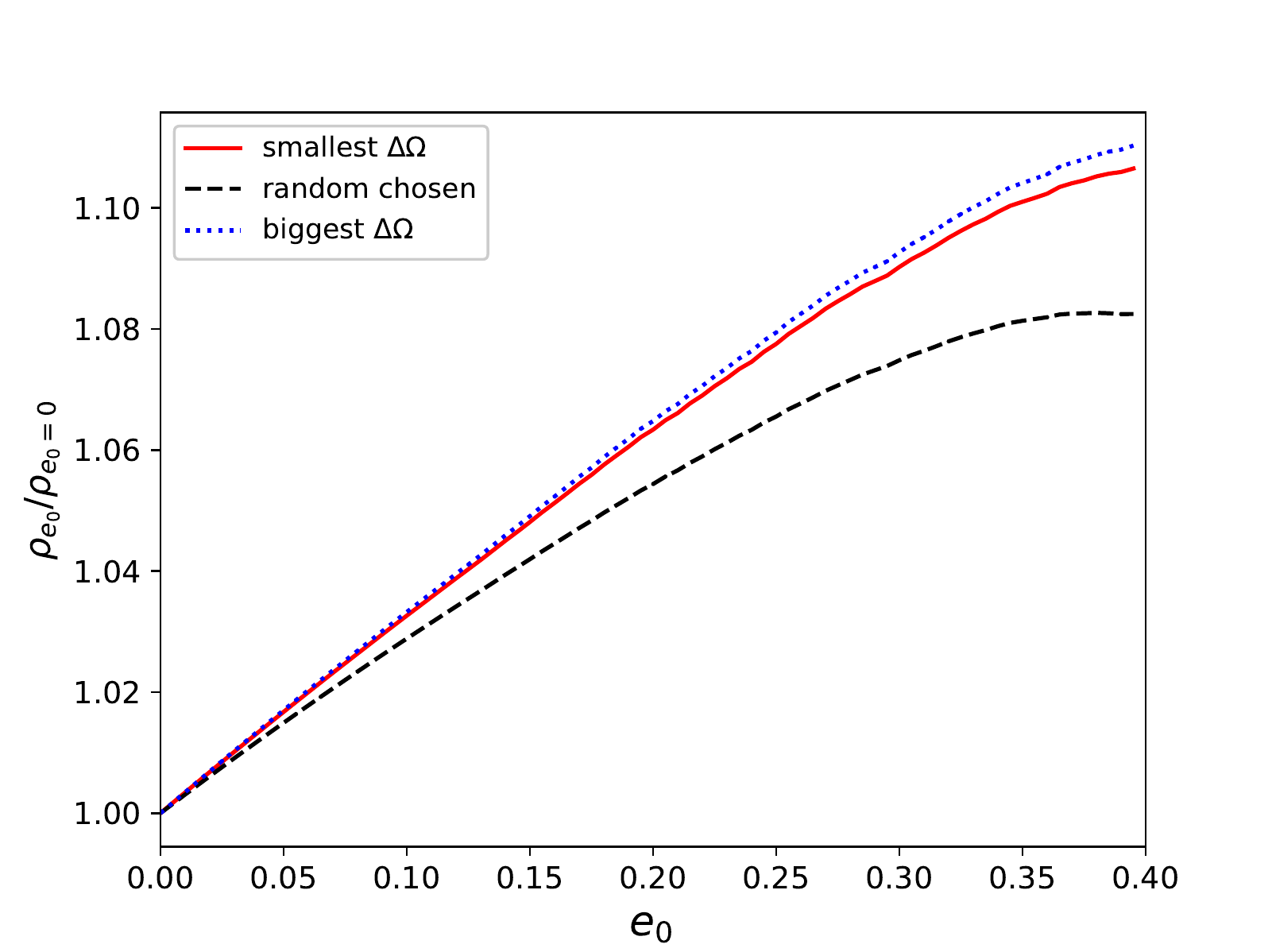}
\end{tabular}
\caption{The improvement behavior of SNR respect to the eccentricity. The upper plot is for big black hole case with total mass 100M${}_\odot$ which corresponds to the Fig.~\ref{fig:100-omega-e}. The lower plot is for GW150914-like black hole case with total mass 65M${}_\odot$ which corresponds to the Fig.~\ref{fig:65-omega-e}.} \label{fig:SNR-e}
\end{figure}

Recalling to the results we got in \cite{LIGOPSD}, we have found that the eccentricity can improve the parameters measurement. Unlike the chirp mass and mass ratio, whose improvement by the eccentricity are independent of the total mass of the binary, the source localization improved as the total mass increases.
Regarding to this difference ones may suspect that this eccentricity-enhanced improvement in the source localization
may result from that on the SNR. We plot the improvement of SNR respect to the eccentricity in the Fig.~\ref{fig:SNR-e}. The two panels are corresponding to the Figs.~\ref{fig:100-omega-e} and \ref{fig:65-omega-e} respectively. The improvement of SNR respect to the eccentricity depends on the total mass of the binary can thus be easily understood. As shown by previous studies, the eccentricity may excite higher frequency signal compared to the circular system. Within EPC model, we can see that the waveform frequency may reach $\ell$F${}_\text{LSO}$ for $\ell$ mode. When the eccentricity is as large as 0.4, the $\ell=10$ mode should be considered. So the waveform will reach 10F${}_\text{LSO}$. Compared to the circular corresponding system, the frequency range 2F${}_\text{LSO}<f<$ 10F${}_\text{LSO}$ is the bonus introduced by the eccentricity. For total mass 100M${}_\odot$ big binary black hole system, this frequency range is 44Hz $<f<$ 220Hz which falls in the most sensitive part of the AdvLIGO band. For total mass 65M${}_\odot$ GW150914-like binary black hole system, this frequency range is 68Hz $<f<$ 340Hz which falls in the AdvLIGO band but less sensitive part. For total mass 22M${}_\odot$ GW151226-like binary black hole system, this frequency range is 100Hz $<f<$ 500Hz which falls in the much less sensitive part of AdvLIGO band. So the SNR improves stronger by the eccentricity for big binary black hole case than for GW150914-like case as shown in the Fig.~\ref{fig:SNR-e}. When the total mass decreases to about 20M${}_\odot$, the SNR does not depend on the eccentricity anymore. We have seen this result in \cite{LIGOPSD} also.

\begin{figure}
\begin{tabular}{c}
\includegraphics[width=0.48\textwidth]{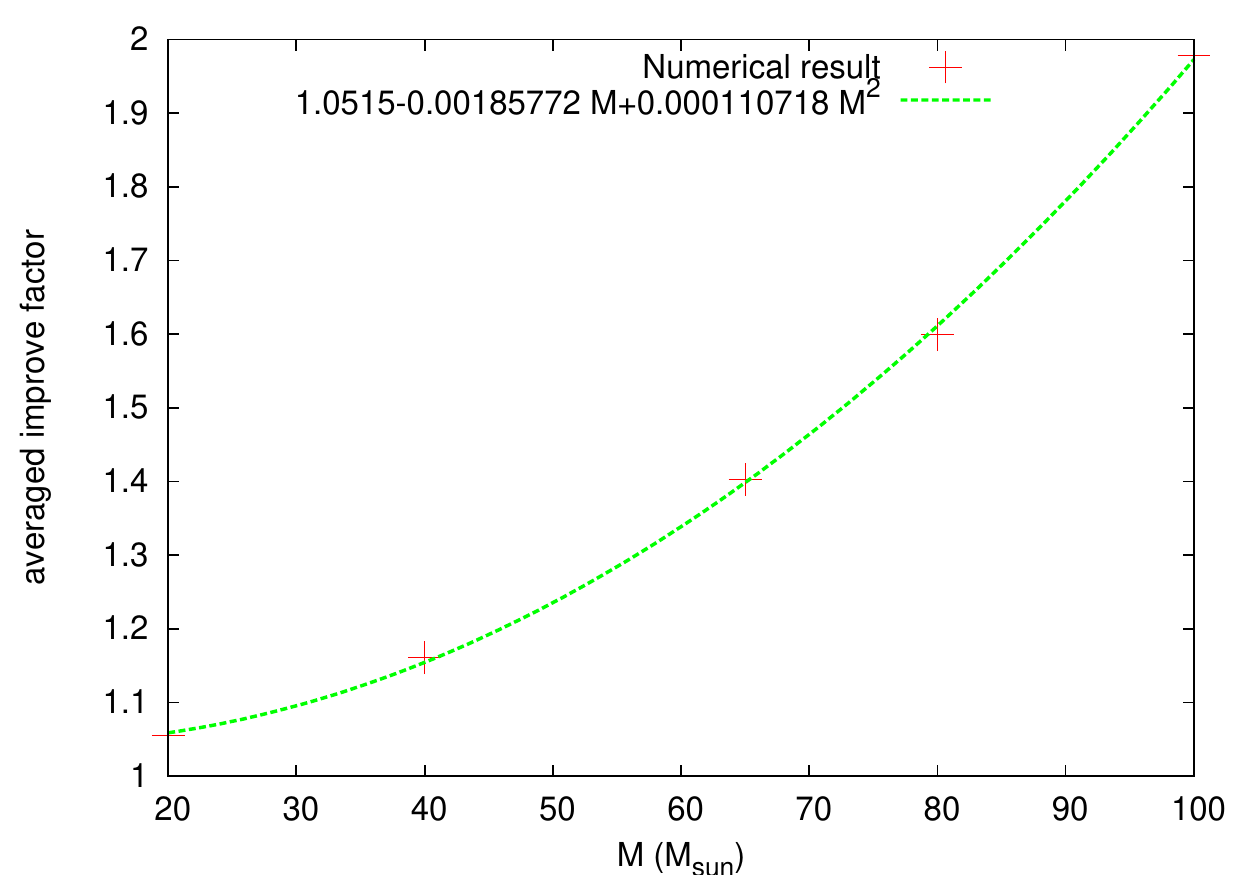}
\end{tabular}
\caption{The averaged improvement factor $\frac{1}{4\pi}\int\frac{\Delta\Omega|_{e_0=0.4}}{\Delta\Omega|_{e_0}}\sin\theta_ed\theta_ed\phi_e$ for binary black hole systems with different total masses. The dotted line is the quadratic fitting result.} \label{fig13}
\end{figure}

Comparing the Figs.~\ref{fig:100-omega-e} and \ref{fig:65-omega-e} to the Fig.~\ref{fig:SNR-e}, we find that the source localization improvement depends not only on the improvement of the SNR but also on the structure of the eccentric waveform. In order to investigate the dependence of the source localization improvement by the eccentricity on the the total mass of the binary, we calculate the improvement factor between $e=0$ and $e=0.4$ systematically. Besides the above reported binary systems with total mass 20M${}_\odot$, 65M${}_\odot$, and 100M${}_\odot$, we consider more binary systems with total mass 40M${}_\odot$ and 80M${}_\odot$. For each binary system, we chose parameters $D_{Le}=410$Mpc, $\eta=0.25$, $t_{ce}=0$, $\phi_{c}=0$, $\iota_{e}=0$, $\beta_{e}=0$, and $\psi_{e}=0$ as example, while survey all $(\theta_e, \phi_e)$ parameters space. Then we average the improvement factor between $e=0$ and $e=0.4$ respect to the whole $(\theta_e, \phi_e)$ parameters space. We plot the result in the Fig.~\ref{fig13}. Along with the numerical calculation results, we also plot the quadratic fitting result in this figure. Overall, we can see that the source localization improvement depends on the total mass of the binary system quadratically.

\begin{acknowledgments}
This work was supported by the NSFC (No.~11690023, No.~11622546, and No.~11375260). Z Cao was supported by ``the Fundamental Research Funds for the Central Universities".
\end{acknowledgments}

\appendix

\section{Relationship between the earth center based coordinate and the detector based coordinate}\label{app:A}
\begin{figure}
\begin{tabular}{c}
\includegraphics[width=0.5\textwidth]{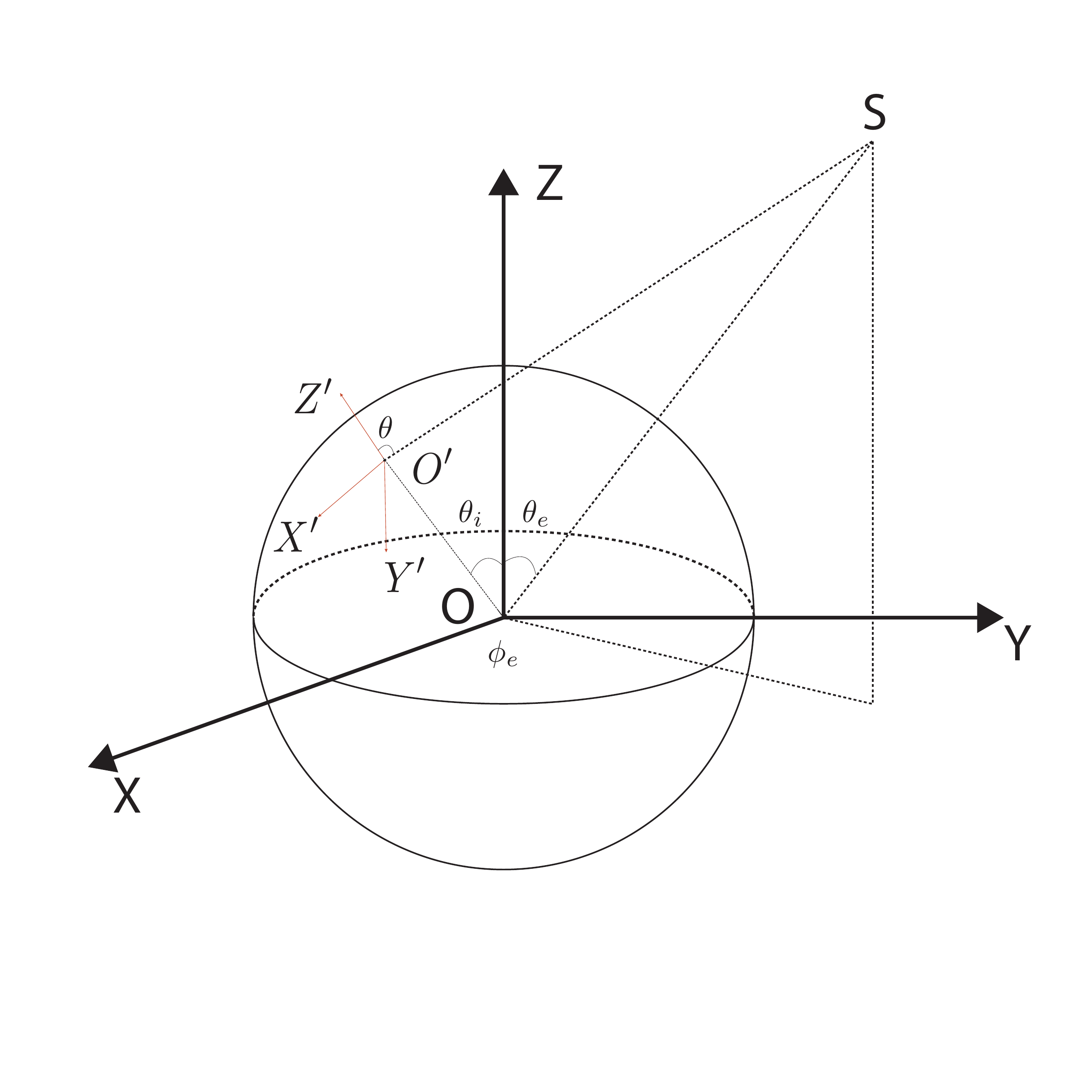}
\end{tabular}
\caption{The earth-based coordinate $O\text{-}XYZ$ and the  detector based coordinate $\text{O}^{\prime}\text{-}\text{X}^{\prime}\text{Y}^{\prime}\text{Z}^{\prime}$. The sphere represents the Earth. $S$ is the gravitational wave source and $O^{\prime}$ is the location of the detector.}\label{fig:theta}
\end{figure}

In this appendix, we deduce the relationship between the earth center based coordinate and the detector based coordinate. As shown in Fig.~\ref{fig:theta}, the sphere represents the Earth, and S denotes the gravitational wave source. $O^{\prime}$ corresponds to the location of detector. As mentioned in the Sec.~\ref{sec:EPC}, we assume the location of detector are $(\theta_{i},\phi_{i})$, where $\theta_{i}$ and $\phi_{i}$ are colatitude and longitude of $i$-th detector. Or to say they are respect to the earth based coordinate. Then we have the relation
\begin{align}
D^2_{L}&=(D_{Le}\sin\theta_{e}\cos\phi_e-R_e\sin\theta_i\cos\phi_i)^2\notag\\
&+(D_{Le}\sin\theta_{e}\sin\phi_e
-R_e\sin\theta_i\sin\phi_i)^2\notag\\
&+(D_{Le}\cos\theta_{e}-R_e\cos\theta_i)^2,\label{DEr_D}\\
t_c&=t_{ce}+D_{L}-D_{Le}.\label{DEr_t}
\end{align}
We use $O\text{-}XYZ$ to denote the Earth-based coordinate, where $OX$ lies on the equatorial plane and points to longitude 0 direction. $OZ$ points to the north direction. We use $\hat{e}_x$, $\hat{e}_y$ and $\hat{e}_z$ to denote the coordinate basis vectors. Regarding the detector-based coordinate $O^{\prime}\text{-}X^{\prime}Y^{\prime}Z^{\prime}$, $O^{\prime}X^{\prime}$ and $O^{\prime}Y^{\prime}$ are along the two arms of detector, $O^{\prime}Z^{\prime}$ is pointing out of the earth surface. We use $\hat{e}'_x$, $\hat{e}'_y$ and $\hat{e}'_z$ to denote the corresponding coordinate basis vectors. The x-arm of the detector ($\hat{e}'_x$) rotates an angle $\psi_i$ from north direction to west direction. Consequently, the y-arm of the detector ($\hat{e}'_y$) rotates an angle $\psi_i$ from west to south.

For convenience, we introduce two intermediate vectors $\hat{e}_n$ and $\hat{e}_w$ which are along north and west directions, respectively. Hence we have
\begin{equation}
\begin{split}
&\hat{e}_n=-\cos\theta_{i}\cos\phi_{i}\hat{e}_x-\cos\theta_{i}\sin\phi_{i}\hat{e}_y+\sin\theta_{i}\hat{e}_z, \\
&\hat{e}_w=\sin\phi_{i}\hat{e}_x-\cos\phi_{i}\hat{e}_y, \\
&\hat{e}'_x=\cos\psi_i\hat{e}_n+\sin\psi_i\hat{e}_w,\\
&\hat{e}'_y=-\sin\psi_i\hat{e}_n+\cos\psi_i\hat{e}_w.
\end{split}
\end{equation}
So we get
\begin{align}
\hat{e}'_x&=[-\cos\psi_i\cos\theta_i\cos\phi_i+\sin\psi_i\sin\phi_i]\hat{e}_x\notag\\
  &-(\cos\psi_i\cos\theta_i\sin\phi_i+\sin\psi_i\cos\phi_i)\hat{e}_y\notag\\
  &+\sin\theta_i\cos\psi_i\hat{e}_z\label{eqepx}\\
\hat{e}'_y&=[\sin\psi_i\cos\theta_i\cos\phi_i+\cos\psi_i\sin\phi_i]\hat{e}_x\notag\\
  &+[\sin\psi_i\cos\theta_i\sin\phi_i-\cos\psi_i\cos\phi_i]\hat{e}_y\notag\\
  &-\sin\theta_i\sin\psi_i\hat{e}_x\label{eqepy}\\
\hat{e}'_z&=\sin\theta_i\cos\phi_i\hat{e}_x+\sin\theta_i\sin\phi_i\hat{e}_y+\cos\theta_i\hat{e}_z.\label{eqepz}
\end{align}
\begin{figure}
\begin{tabular}{c}
\includegraphics[width=0.5\textwidth]{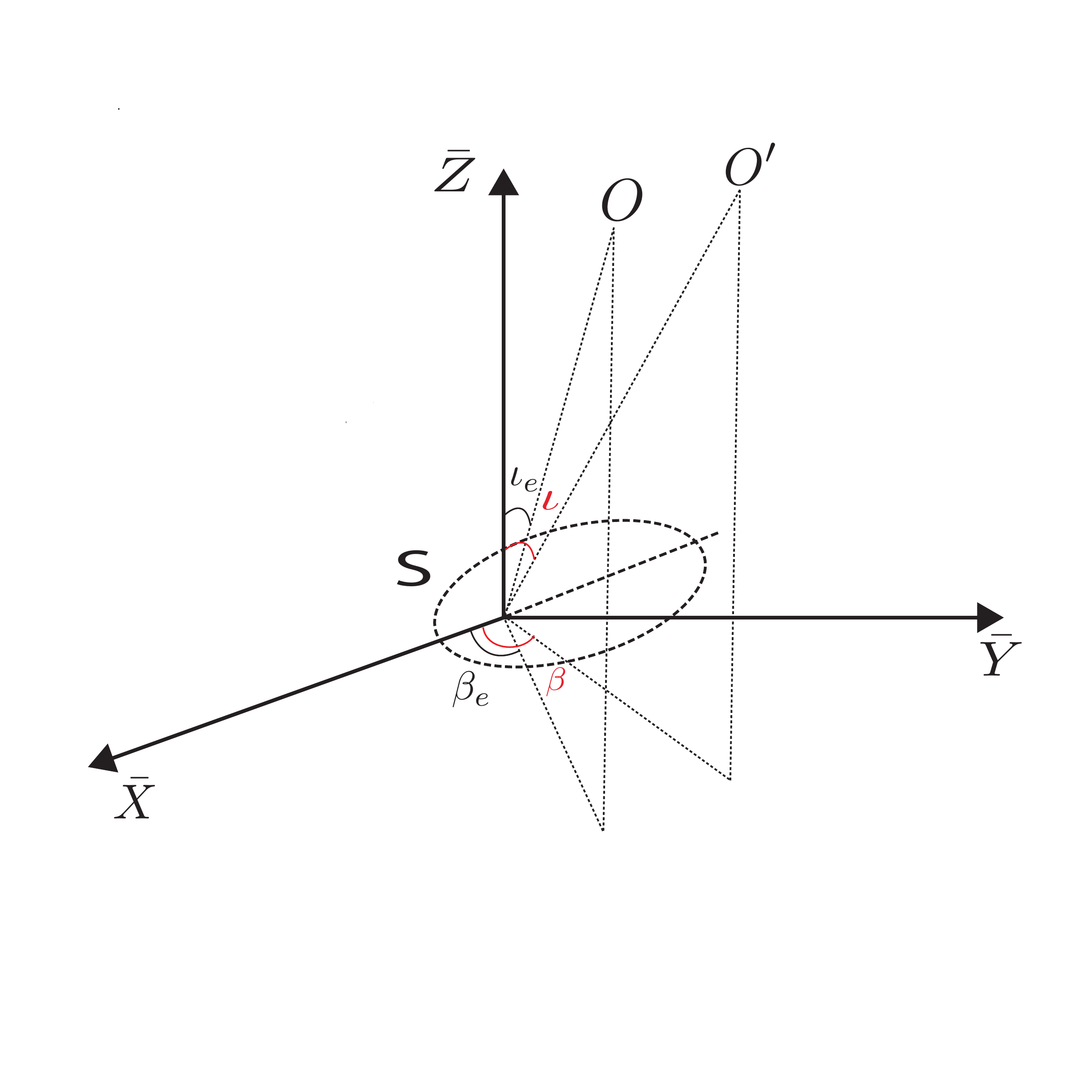}
\end{tabular}
\caption{$S\text{-}\bar{X}\bar{Y}\bar{Z}$ is the source-based coordinate, where $\bar{X}\bar{Y}$ plane is the orbital plane of the binary and $S\bar{Z}$ points along the orbital angular momentum direction. $S$ is the gravitational wave source. $O$ is the center of the Earth. $O'$ is the detector. $OS$ and $O'S$ are theline-of-sight from the earth and the detector, respectively.}\label{fig:iota-beta}
\end{figure}

In Fig.~\ref{fig:iota-beta}, we denote the gravitational wave source with $S$, the center of the Earth with $O$, and the detector with $O'$ as in Fig.~\ref{fig:theta}. So $OS$ and $O'S$ are the sight directions of the source from the earth and the detector respectively. We have constructed the gravitational wave source based coordinate $S\text{-}\bar{X}\bar{Y}\bar{Z}$ for convenience. In this coordinate, $\bar{X}\bar{Y}$ plane coincides with the orbital plane of the binary, $S$ corresponds to the focus of the elliptical orbit, $\overrightarrow{S\bar{Z}}$ points along the orbital angular momentum direction and $\overrightarrow{S\bar{X}}$ points to the periastron direction of the orbit. Respect to the coordinate $S\text{-}\bar{X}\bar{Y}\bar{Z}$ the direction of $\overrightarrow{SO}$ is $(\iota_e,\beta_e)$. So if we rotate the coordinate $S\text{-}\bar{X}\bar{Y}\bar{Z}$ along $S\text{-}\bar{Z}$ axis $\beta_e$ firstly and along the new $S\text{-}\bar{Y}$ axis $\iota_e$ then, the new $S\text{-}\bar{Z}$ axis will be parallel to the direction $\overrightarrow{SO}$. On the other hand, the direction of $\overrightarrow{OS}$ is $(\theta_e,\phi_e)$ respect to the coordinate $O\text{-}XYZ$. Or to say the direction of $\overrightarrow{SO}$ is $(\pi-\theta_e,\phi_e)$. So if we rotate the coordinate $O\text{-}XYZ$ along $O\text{-}Z$ axis $\phi_e$ firstly and along the new $O\text{-}Y$ axis $\pi-\theta_e$ then, the new $O\text{-}Z$ axis will be parallel to the direction $\overrightarrow{SO}$. Now the new $O\text{-}X$ axis will not be parallel to the new $S\text{-}\bar{X}$ axis. This angle is nothing but the usual called polarization angle $\psi_e$. So we can rotate the new coordinate $S\text{-}\bar{X}\bar{Y}\bar{Z}$ along $S\text{-}\bar{Z}$ axis $\psi_e$ to make $S\text{-}\bar{X}\bar{Y}\bar{Z}$ parallel to $O\text{-}XYZ$ axis by axis. Equivalently we have
\begin{align}
&\left(\begin{matrix}
\cos\psi_e  & \sin\psi_e & 0 \\
-\sin\psi_e & \cos\psi_e & 0 \\
0 & 0 & 1
\end{matrix}
\right)
\left(\begin{matrix}
\cos\iota_e  & 0 & -\sin\iota_e\\
0 & 1 & 0\\
\sin\iota_e & 0 & \cos\iota_e
\end{matrix}
\right)
\times\notag\\
&
\left(\begin{matrix}
\cos\beta_e  & \sin\beta_e & 0 \\
-\sin\beta_e & \cos\beta_e & 0 \\
0 & 0 & 1
\end{matrix}
\right)
\left(
\begin{matrix}
\hat{e}_{\bar{x}} \\
\hat{e}_{\bar{y}} \\
\hat{e}_{\bar{z}}
\end{matrix}
\right)=\nonumber\\
&\left(\begin{matrix}
\cos(\pi-\theta_e) & 0 & -\sin(\pi-\theta_e)\\
0 & 1  & 0\\
\sin(\pi-\theta_e) & 0 & \cos(\pi-\theta_e)
\end{matrix}
\right)\times\nonumber\\
&
\left(\begin{matrix}
\cos\phi_e  & \sin\phi_e & 0 \\
-\sin\phi_e & \cos\phi_e & 0 \\
0 & 0 & 1
\end{matrix}
\right)
\left(
\begin{matrix}
\hat{e}_{x} \\
\hat{e}_{y} \\
\hat{e}_{z}
\end{matrix}
\right).\label{reebepxyz}
\end{align}
And more we can decompose $\overrightarrow{SO'}$ respect to the $S\text{-}\bar{X}\bar{Y}\bar{Z}$ coordinate basis or the $O'\text{-}X'Y'Z'$ coordinate basis as
\begin{align}
\overrightarrow{SO'}&=D_L[\sin\iota\cos\beta\hat{e}_{\bar{x}}+\sin\iota\sin\beta\hat{e}_{\bar{y}}
+\cos\iota\hat{e}_{\bar{z}}]\label{sopeb}\\
&=\overrightarrow{SO}+\overrightarrow{OO^\prime} \\
&=D_{Le}[\sin\iota_e\cos\beta_e\hat{e}_{\bar{x}}
+\sin\iota_e\sin\beta_e\hat{e}_{\bar{y}}+\cos\iota_e\hat{e}_{\bar{z}}]\notag\\
&+R_e[\sin\theta_i\cos\phi_i\hat{e}_{x}+\sin\theta_i\sin\phi_i\hat{e}_{y}+\cos\theta_i\hat{e}_{z}].\label{sopep}
\end{align}
We have used $D_L$ to denote the length of $\overrightarrow{SO'}$. Then plugging Eq.~(\ref{reebepxyz}) into Eq.~(\ref{sopep}) we can get
\begin{align}
\cos\iota&=\frac{D_{Le}}{D_{L}}\cos\iota_e
+\frac{R_e}{D_L}\{\cos\theta_e[\cos\theta_i\cos\iota_e\notag\\
&-\sin\theta_i\sin\iota_e\cos\psi_e\cos(\phi_e-\phi_i)]\notag \\
&+\sin\iota_e[\sin \theta_e\cos\theta_i\cos\psi_e-\sin\theta_i\sin\psi_e\sin(\phi_e-\phi_i)]\notag\\
&+\sin\theta_e\sin\theta_i\cos\iota_e\cos(\phi_e-\phi_i)\},\label{DEr_iota}\\
\tan\beta&=\{D_{Le}\sin\beta_e\sin\iota_e+R_e[-\cos\beta_e\cos\theta_i\sin\psi_e\sin\theta_e\notag\\
&+\cos\beta_e\sin\theta_i\cos\theta_e\cos(\phi_e-\phi_i)\sin\psi_e\notag \\
&-\cos\beta_e\sin\theta_i\cos\psi_e\sin(\phi_e-\phi_i)\notag\\
&-\sin\beta_e\cos\psi_e\cos\iota_e\cos\theta_i\sin\theta_e\notag\\
&+\sin\beta_e\cos\psi_e\cos\iota_e\cos\theta_e\cos(\phi_e-\phi_i)\sin\theta_i\notag \\
&+\sin\beta_e\cos\theta_e\cos\theta_i\sin\iota_e\notag\\
&+\sin\beta_e\sin\theta_i\cos\phi_e\cos\phi_i\sin\theta_e\sin\iota_e\notag\\
&-\sin\beta_e\sin\theta_i\cos\phi_e\cos\iota_e\sin\psi_e\sin\phi_i\notag \\
&+\sin\beta_e\sin\theta_i\sin\phi_e\cos\iota_e\cos\phi_i\sin\psi_e\notag\\
&+\sin\beta_e\sin\theta_i\sin\phi_e\sin\theta_e\sin\iota_e\sin\phi_i]\}\notag \\
&/\{D_{Le}\cos\beta_e\sin\iota_e\notag\\
&+R_e[\cos\beta_e\cos\theta_e \sin\theta_i\cos\iota_e\cos\psi_e\cos(\phi_e-\phi_i)\notag\\
&-\cos\beta_e\sin\theta_e\cos\theta_i\cos\iota_e\cos\psi_e\notag \\
&+\cos\beta_e\sin\theta_e\sin\theta_i\sin\iota_e\cos\phi_e\cos \phi_i\notag\\
&+\cos\beta_e\sin\theta_e\sin\theta_i\sin\iota_e\sin\phi_e \sin\phi_i\notag\\
&+\cos\beta_e\cos\theta_e\cos\theta_i\sin\iota_e\notag \\
&-\sin\beta_e\cos\theta_e\sin\theta_i\sin\psi_e\cos(\phi_e-\phi_i)\notag\\
&+\sin\beta_e\sin\theta_e\cos\theta_i\sin\psi_e\notag\\
&+\cos\beta_e\sin\theta_i\cos\iota_e\sin\psi_e\sin\phi_e\cos\phi_i\notag \\
&-\cos\beta_e\sin\theta_i\cos\iota_e\sin\psi_e\cos\phi_e\sin\phi_i\notag\\
&+\sin\beta_e\sin\theta_i\cos\psi_e\sin(\phi_e-\phi_i)]\}.\label{DEr_beta}
\end{align}

As shown in Fig.~\ref{fig:theta}, the vector $\overrightarrow{O'S}$ can be expressed as
\begin{equation}
\begin{split}
\overrightarrow{O'S}&=\overrightarrow{OS}-\overrightarrow{OO'}\\
&=(D_{Le}\sin\theta_e\cos\phi_e-R_e\sin\theta_i\cos\phi_i)\hat{e}_x\\
&+(D_{Le}\sin\theta_e\sin\phi_e-R_e\sin\theta_i\sin\phi_i)\hat{e}_y\\
&+(D_{Le}\cos\theta_e-R_e\cos\theta_i)\hat{e}_z,
\end{split}
\end{equation}
where $R_e$ is the radius of the Earth, $D_{Le}$ is the length of $\overrightarrow{OS}$, $(\theta_e,\phi_e)$ is the source angular position respect to the earth based coordinate. the source angular position respect to the detector based coordinate $(\theta,\phi)$ can be expressed as
\begin{align}
&\cos\theta=\frac{\overrightarrow{O'S}}{|O'S|}\cdot\hat{e}'_{z},\\
&\sin\theta\cos\phi=\frac{\overrightarrow{O'S}}{|O'S|}\cdot\hat{e}'_{x},  \\
&\sin\theta\sin\phi=\frac{\overrightarrow{O'S}}{|O'S|}\cdot\hat{e}'_{y}.
\end{align}
Then plugging Eqs.~(\ref{eqepx})-(\ref{eqepz}) into the above equations and straight forward calculation will result in
\begin{align}
\cos\theta&=\frac{D_{Le}}{D_{L}}[\sin\theta_{e}\sin\theta_{i}\cos(\phi_{e}-\phi_{i})
+\cos\theta_{e}\cos\theta_{i}]\notag\\
&-\frac{R_{e}}{D_{L}},\label{DEr_theta}\\
\tan\phi&=[\sin\theta_{e}\sin\psi_i\cos\theta_{i}\cos(\phi_{e}-\phi_{i})\notag\\
&-\cos\theta_{e}\sin\theta_{i}
\sin\psi_i-\sin\theta_{e}\cos\psi_i\sin(\phi_{e}-\phi_{i})] \notag \\
&/[-\sin\theta_{e}\cos\psi_i\cos\theta_{i}\cos(\phi_{e}-\phi_{i})\notag\\
&+\cos\theta_{e}\sin\theta_{i}\cos\psi_i +\sin\theta_{e}\sin\psi_i\sin(\phi_{i}-\phi_{e})].\label{DEr_phi}
\end{align}

\begin{figure}
\begin{tabular}{c}
\includegraphics[width=8.6cm]{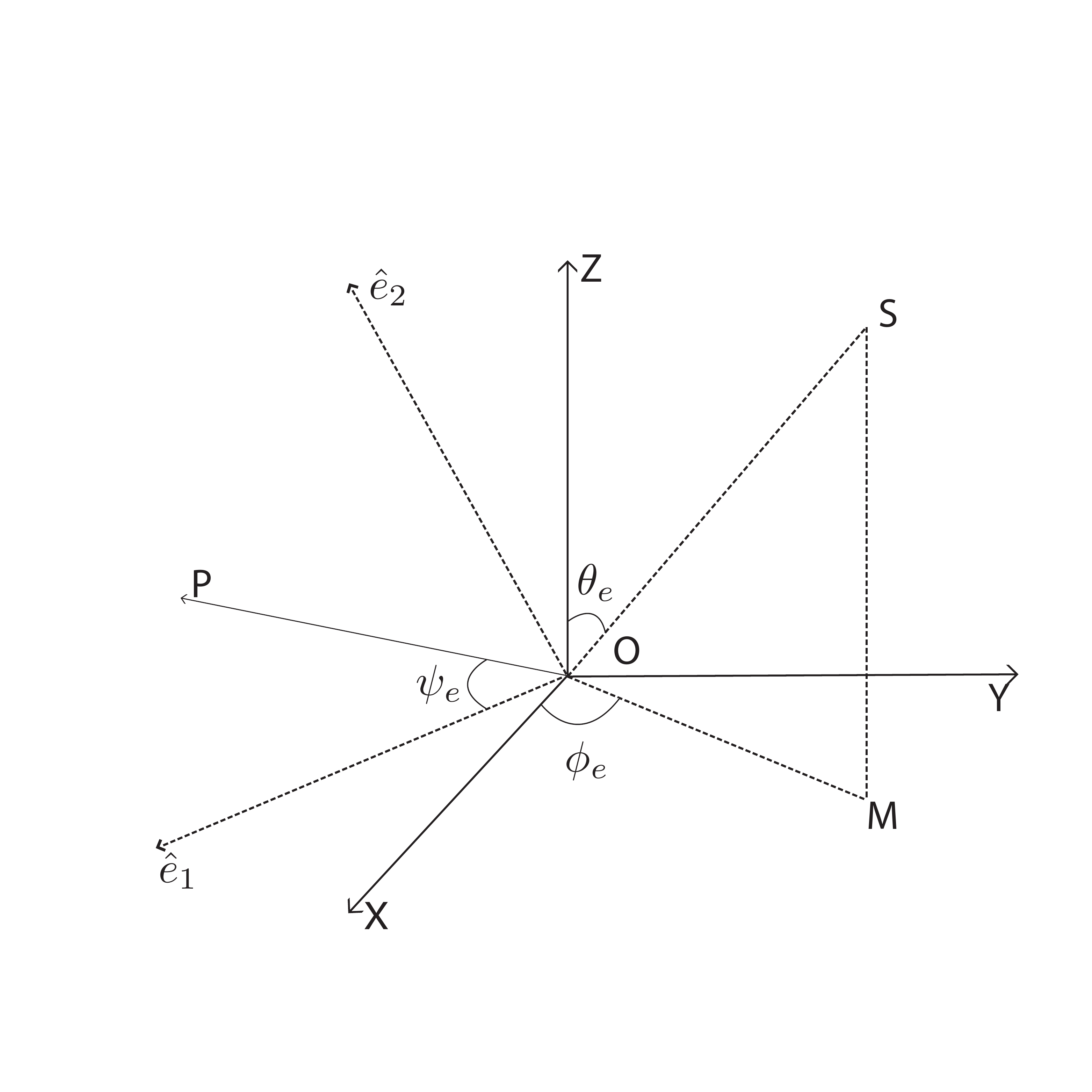}
\end{tabular}
\caption{$O\text{-}XYZ$ is the earth-based coordinate. $S$ is the gravitational wave source. $\hat{e}_{1}$ and $\hat{e}_{2}$ are orthogonal to $\protect\overrightarrow{OS}$. $\hat{e}_{1}$ is on the $O\text{-}XY$ plane. $\protect\overrightarrow{OP}$ is the oscillation vector of the gravitational wave. $\psi$ is the angle between $\protect\overrightarrow{OP}$ and $\hat{e}_{1}$.}
\label{fig:psi}
\end{figure}
Suppose that the oscillation basis vectors of gravitational wave correspond to $\overrightarrow{OP}$ and $\overrightarrow{OQ}$. Due to the transverse property of gravitational wave, the plane determined by $\overrightarrow{OP}$ and $\overrightarrow{OQ}$ is perpendicular to $\overrightarrow{OS}$. Assume the node line between this plane and $XY$ plane is $\hat{e}_{1}$. Then the polarization angle $\psi_e$ is the angle between $\overrightarrow{OP}$ and $\hat{e}_{1}$. For convenience we introduce a $\hat{e}_{2}$ which lies in the oscillation plane of the gravitational wave and perpendicular to $\hat{e}_{1}$. Based on $\hat{e}_{1}$ and $\hat{e}_{2}$ we can decompose the unit vector $\overrightarrow{OP}$ as
\begin{equation}
\overrightarrow{OP}=\cos\psi_e\hat{e}_{1}+\sin\psi_e\hat{e}_{2}.\label{eqop}
\end{equation}
Note $\hat{e}_{1}$ is perpendicular to $OS$ and $SM$, so $\hat{e}_{1}$ is perpendicular to $OM$. Then the angle formed by $\hat{e}_{1}$ and $OX$ is $\frac{\pi}{2}-\phi_e$. Note $\hat{e}_{1}$ is perpendicular to $\hat{e}_{2}$, $OZ$, $OS$ and $OM$, we can deduce $\hat{e}_{2}$, $OZ$, $OS$ and $OM$ lie in the same plane. Based on these facts, we have
\begin{align}
&\hat{e}_{1}=\sin\phi_e\hat{e}_{x}-\cos\phi_e\hat{e}_{y}, \\
&\hat{e}_{2}=-\cos\phi_e\cos\theta_e\hat{e}_{x}-\sin\phi_e\cos\theta_e\hat{e}_{y}+\sin\theta_e\hat{e}_{z}.
\end{align}
Here we neglect the difference between the oscillation basis vectors of gravitational wave at $O$ and $O'$. Then if we replace $O\text{-}XYZ$ with $O'\text{-}X'Y'Z'$ we can construct similarly $\hat{e}'_1$, $\hat{e}'_2$ and the polarization angle $\psi$ respect to the detector coordinate which is the angle between $\overrightarrow{OP}$ and $\hat{e}'_{1}$. And we have the relation
\begin{align}
&\hat{e}'_{1}=\sin\phi\hat{e}'_{x}-\cos\phi\hat{e}'_{y}, \label{eqe1p}\\
&\hat{e}'_{2}=-\cos\phi\cos\theta\hat{e}'_{x}-\sin\phi\cos\theta\hat{e}'_{y}+\sin\theta\hat{e}'_{z}.\label{eqe2p}
\end{align}
According to the definition of $\psi$ we have
\begin{align}
\cos\psi=&\overrightarrow{OP}\cdot\hat{e}'_{1},\\
\sin\psi=&\overrightarrow{OP}\cdot\hat{e}'_{2}.
\end{align}
Then plugging in Eqs.~(\ref{eqop})-(\ref{eqe2p}) and (\ref{eqepx})-(\ref{eqepz}) we can get
\begin{align}
&\cos\psi=\notag\\
&\sin\psi_{e}\sin\theta_{i}\sin(\phi+\psi_i)
\sqrt{1-(\frac{D_{Le}}{D_{L}}\cos\theta_{e}-\frac{R_e}{D_{L}}\cos\theta_{i})^{2}} \notag \\
&-\cos(\phi+\psi_i)\cos\psi_{e}\cos\{\notag\\
&\arctan\frac{D_{Le}\sin\theta_{e}\sin\phi_e-R_e\sin\theta_i\sin\phi_i}
{D_{Le}\sin\theta_{e}\cos\phi_e-R_e\sin\theta_i\cos\phi_i}-\phi_{i}\} \notag\\
&-(\frac{D_{Le}}{D_{L}}\cos\theta_{e}-\frac{R_e}{D_{L}}\cos\theta_{i})\cos(\phi+\psi_i)\sin\psi_{e}\sin\{\notag\\
&\arctan\frac{D_{Le}\sin\theta_{e}\sin\phi_e-R_e\sin\theta_i\sin\phi_i}{
D_{Le}\sin\theta_{e}\cos\phi_e-R_e\sin\theta_i\cos\phi_i}-\phi_{i}\}\notag \\
&-\sin(\phi+\psi_i)\cos\psi_{e}\cos\theta_{i}\sin\{\notag\\
&\arctan\frac{D_{Le}\sin\theta_{e}\sin\phi_e
-R_e\sin\theta_i\sin\phi_i}{D_{Le}\sin\theta_{e}\cos\phi_e-R_e\sin\theta_i\cos\phi_i}-\phi_{i}\}\notag \\
&+(\frac{D_{Le}}{D_{L}}\cos\theta_{e}-\frac{R_e}{D_{L}}\cos\theta_{i})\sin(\phi+\psi_i)
\sin\psi_{e}\cos\theta_{i}\cos\{\notag\\
&\arctan\frac{D_{Le}\sin\theta_{e}\sin\phi_e
-R_e\sin\theta_i\sin\phi_i}{D_{Le}\sin\theta_{e}\cos\phi_e-R_e\sin\theta_i\cos\phi_i}-\phi_{i}\}.\label{DEr_psi}
\end{align}
Till now we have got the complete relationships between the earth center based coordinate and the detector based coordinate including Eqs. (\ref{DEr_D}), (\ref{DEr_t}), (\ref{DEr_iota}), (\ref{DEr_beta}), (\ref{DEr_theta}), (\ref{DEr_phi}) and (\ref{DEr_psi}).

\bibliographystyle{apsrev}
\bibliography{refer}
\end{document}